\pdfoutput=1

\documentclass[11pt,twoside,a4paper,cmspaper,final,collab]{cms-tdr}
\def\svnVersion{290990}\def\cmsCernNo{2013-037}\def\cmsCernDate{\today}\def\cmsMessage{Published in Physical Review D as \href{http://dx.doi.org/10.1103/PhyRevD.91.092012}{\doi{10.1103/PhyRevD.91.092012.}}}

\begin{document}\cmsNoteHeader{EWK-11-016}

\hyphenation{had-ron-i-za-tion}
\hyphenation{cal-or-i-me-ter}
\hyphenation{de-vices}
\RCS$Revision: 118704 $
\RCS$HeadURL: svn+ssh://svn.cern.ch/reps/tdr2/papers/EWK-11-016/trunk/EWK-11-016.tex $
\RCS$Id: EWK-11-016.tex 118704 2012-04-27 16:21:30Z akub19 $
\ifthenelse{\boolean{cms@external}}{\providecommand{\cmsLeft}{top\xspace}}{\providecommand{\cmsLeft}{left\xspace}}
\ifthenelse{\boolean{cms@external}}{\providecommand{\cmsRight}{bottom\xspace}}{\providecommand{\cmsRight}{right\xspace}}
\ifthenelse{\boolean{cms@external}}{\newcommand{\cmsTable}[2]{\relax#2}}{\newcommand{\cmsTable}[2]{\resizebox{#1}{!}{#2}}}

\newcommand{\DRmg}    {\ensuremath{\Delta R_{\mu\gamma}}\xspace}
\newcommand{\Imu}     {\ensuremath{I_{\mu}}\xspace}
\newcommand{\Igamma}  {\ensuremath{I_{\gamma}}\xspace}
\newcommand{\DXDET}   {\ensuremath{\rd \sigma / \rd \ET}\xspace}
\newcommand{\DXDDR}   {\ensuremath{\rd \sigma / \rd \Delta R_{\mu\gamma}}\xspace}
\newcommand{\mpmm}    {\ensuremath{\Pgmp\Pgmm}\xspace}
\newcommand{\Mmm}     {\ensuremath{M_{\Pgm\Pgm}}\xspace}
\newcommand{\Mmmg}    {\ensuremath{M_{\Pgm\Pgm\Pgg}}\xspace}
\newcommand{\EgammaPF} {\ensuremath{E_{\gamma}^{\mathrm{PF}}}\xspace}
\newcommand{\EgammaKIN}{\ensuremath{E_{\gamma}^{\text{kin}}}\xspace}
\newcommand{\ETgammaPF} {\ensuremath{E_{{\mathrm{T}}\gamma}^{\mathrm{PF}}}\xspace}
\newcommand{\ETgammaKIN}{\ensuremath{E_{{\mathrm{T}}\gamma}^{\text{kin}}}\xspace}
\newcommand{\QT}      {\ensuremath{q_{\mathrm{T}}}\xspace}
\newcommand{\THELUMI} {4.7\fbinv}
\newcommand{\NSIGNAL} {56\,005\xspace}
\newcommand{\NCONTROL}{45\,277\xspace}
\newcommand{\FRACSIGNALMUMU} {18.3\%\xspace}
\newcommand{\FRACCLOSE} {\ensuremath{8.7 \pm 0.1\stat\pm 0.2\syst\%}\xspace}
\newcommand{\FRACWIDE}  {\ensuremath{5.6 \pm 0.1\stat\pm 0.2\syst\%}\xspace}
\newcommand{\FRACRARE}  {\ensuremath{( 1.3 \pm 0.5\stat\pm 0.6\syst) \times 10^{-4}}\xspace}

\cmsNoteHeader{EWK-11-016}
\title{\texorpdfstring{Study of final-state radiation in decays of \Z bosons produced in $\Pp\Pp$ collisions at 7\TeV}{Study of final-state radiation in decays of Z bosons produced in pp collisions at 7 TeV}}

\date{\today}

\abstract{
The differential cross sections for the production of photons in $\Z \to \Pgmp\Pgmm\gamma$ decays
are presented as a function of the transverse energy of the photon
and its separation from the nearest muon.  The data for these measurements were
collected with the CMS detector and correspond to an integrated luminosity
of 4.7\fbinv of $\Pp\Pp$ collisions at $\sqrt{s} = 7$\TeV
delivered by the CERN LHC.   The cross sections are compared to simulations
with {\POWHEG} and  {\PYTHIA}, where {\PYTHIA} is used to simulate parton showers and
final-state photons.  These simulations match the data to better than 5\%.
}

\hypersetup{%
pdfauthor={CMS Collaboration},%
pdftitle={Study of final-state radiation in decays of Z bosons produced in pp collisions at 7 TeV},%
pdfsubject={CMS},%
pdfkeywords={CMS, physics, EWK}}
\maketitle

\section{Introduction}

We present a study and differential cross section measurements of
photons emitted in decays of \Z bosons produced
at a hadron collider.
Such radiative decays of the Z boson were noted in the very first \Z boson publications of UA1 and UA2~\cite{UA1,UA2},
but subsequently have not been given a detailed study in hadron colliders.
In 2011, the CERN LHC delivered
pp~collisions at $\sqrt{s} = 7\TeV$, and data
corresponding to an integrated luminosity of~$\THELUMI$
were collected with the CMS detector.
From these data, we select a sample of events in which a \Z~boson decays to a $\mpmm$ pair and an energetic photon.
We measure the differential cross sections
$\DXDET$ with respect to the photon transverse energy~$\ET$
and $\DXDDR$ with respect to the separation of the photon from the
nearest muon.
   Here,
$ \DRmg = \sqrt{ (\phi_\mu-\phi_\gamma)^2 + (\eta_\mu-\eta_\gamma)^2 } $,
where $\phi$ is the azimuthal angle (in radians) around the beam axis and
$\eta$ is the pseudorapidity.
The cross sections include contributions
from the \Z resonance, virtual photon exchange, and their interference, collectively
referred to as Drell--Yan (DY) production.

Photons emitted in \Z boson decays, which we call final state radiation (FSR)
photons, can be energetic (tens of \GeVns{})
and well separated from the leptons (by more than a radian).
Quantum electrodynamics (QED) corrections that describe FSR production are
well understood.  Quantum chromodynamics (QCD) corrections
modify the kinematic distributions of the \Z boson;
in particular, the \Z boson aquires a nonzero component
of momentum transverse to the beam: $\QT > 0$.
The \FEWZ program calculates both QCD and QED corrections
for the DY process~\cite{bib:FEWZ2.1}.
However, it does not include mixed QCD and QED corrections;
the required two-loop integrals are technically very challenging,
and progress has been made only recently~\cite{bib:Dittmaier}.
In practice, event generators employing matrix element calculations
matched to parton showers must be used~\cite{Bernaciak,Barze}.
It is the goal of this analysis to establish the quality of the modeling of
FSR over a wide kinematic and angular range.
The results will support future
measurements of the \PW~mass, the study of \Z+$\gamma$ production,
and searches for new particles in final states with photons.

In an attempt to compare photons emitted close to a muon
(a process that is  modeled primarily by a partonic photon shower) and far
from the muons (which requires a matrix element calculation),
we measure $\DXDET$ for $0.05 < \DRmg \leq 0.5$ and $0.5 < \DRmg \leq 3$.
Furthermore, since the size of the QCD corrections varies with the transverse momentum
of the Z boson, we measure $\DXDET$
and $\DXDDR$ for $\QT < 10\GeV$ and $\QT > 50\GeV$, where
$\QT$ is defined as the vector sum of the transverse momenta of the two muons and the
photon.  These cross
sections are defined with respect to the fiducial and kinematic
requirements detailed below; no acceptance corrections are applied.
Nonetheless, we do correct for detector resolution and efficiencies.

This article is structured as follows.  We briefly describe the
CMS detector and the event samples we use in Section~\ref{sec:detector}.  The details
of the event selection are given in Section~\ref{sec:selection}. Background estimation
and the way we unfold the data distributions are discussed in Sections~\ref{sec:backgrounds}~and~\ref{sec:unfold}. We discuss
the systematic uncertainties in Section~\ref{sec:systematics} and report our results and
summarize the work in Sections~\ref{sec:results}~and~\ref{sec:summary}.

\section{The CMS detector and event samples}
\label{sec:detector}

A full description of the CMS detector can be found in Ref.~\cite{ref:CMS};
here we briefly describe the components most important for this analysis.
The central feature of the CMS experiment is a superconducting
solenoid that provides an axial magnetic field of 3.8\unit{T}.
A tracking system composed of a
silicon pixel detector and a silicon strip detector is installed around the
beam line, and provides measurements of the trajectory of charged
particles for $\abs{\eta}< 2.5$.
After passing through the tracker, particles strike the crystal
electromagnetic calorimeter (ECAL) followed by the brass and scintillator
hadron calorimeter.  The solenoid coil encloses the tracker and the
calorimetry.  Four stations of muon detectors measure the trajectories
of muons that pass through the tracker and the calorimeters for
$\abs{\eta}< 2.4$.  Three detector technologies are employed in the muon system:
drift tubes for central rapidities, cathode strip detectors for
the forward rapidities, and resistive-plate chambers for all
rapidities.  Combining information from the muon detectors and
the tracker, the transverse momentum~(\PT) resolution for muons
used in this analysis varies from 1 to 6\%, depending on $\eta$
and~\PT~\cite{MUONPAPER}.
The \ET of photons and electrons is measured using
energy deposited in the ECAL, which consists of nearly $76\,000$
lead tungstate crystals distributed in the barrel region
($\abs{\eta}< 1.479$) and two endcap regions ($1.479 < \abs{\eta}< 3.0$).
The photon energy resolution is better than 5\% for the range
of \ET pertinent to this analysis~\cite{EGAMMA}.  Events are selected
using a two-level trigger system.  The level-1 trigger, composed of
custom-designed processing hardware, selects events of interest
based on information from the muon detectors and calorimeters~\cite{L1TDR}.
The high-level trigger is software-based, running a simpler and therefore faster version of
the offline reconstruction code on the full detector information,
including the tracker~\cite{HLTTDR}.

Simulated data samples are used to design and verify the
principles of the analysis.  They are also used to assess
efficiencies, resolution, and backgrounds.  The signal
process is simulated using the \POWHEG~(V1.0)~\cite{POWHEG} event generator with
\PYTHIA~(V6.4.24)~\cite{Pythia6} used to simulate parton showers and
final state photons (referred to in what follows
as \POWHEG{}+\PYTHIA).  This combination is also
used for $\ttbar$ and diboson ($\PW\PW$, $\PW\Z$, $\Z\Z$) production.
The CT10~\cite{CT10} parton distribution functions are used.
The Z2 parameter set~\cite{Z2} is used to model the
underlying event in \PYTHIA, and
the effects of additional $\Pp\Pp$ collisions that produce
signals registered together with the main interaction
are included in the simulation.

The response of the detector is simulated using \GEANTfour~\cite{GEANT4}.
The simulated events are processed using the same version of
the offline reconstruction code used for the data.

\section{Event selection}
\label{sec:selection}

The data are recorded using a trigger that requires two muons.
One muon is required to have $\PT > 13\GeV$,
and the other, $\PT > 8\GeV$.  This trigger has no requirement on the
isolation of the muons.

Events with a pair of oppositely charged,
well-reconstructed, and isolated muons and an isolated
photon are selected.  The kinematic and fiducial requirements for
selecting events are based wholly on the muon and photon kinematic
quantities, and are
summarized in Table~\ref{tab:cuts}.   As explained below,
we use the dimuon mass $\Mmm$ to define a ``signal
region'' that is rich in FSR photons, and a ``control
region'' that is dominated by background sources of photons.

Muons are selected in the manner developed for the
measurements of the DY cross section~\cite{DY}.
They must be reconstructed using an algorithm that finds a
track segment in the muon detectors and links it with a
track in the silicon tracker, and also through an algorithm that extrapolates
a track in the silicon tracker outward and matches it with hits registered
in the muon detectors.
We select the two highest \PT muons (which we will call
``leading'' and ``trailing''), and ignore any additional muons.
These two muons must have opposite charge.
The leading muon must satisfy the requirements
$\PT > 31\GeV$ and $\abs{\eta}< 2.4$, while the trailing
muon must satisfy $\PT > 9\GeV$ and $\abs{\eta}< 2.4$
to ensure good reconstruction efficiency.
A vertex fit is performed to the two muon tracks, and the
$\chi^{2}$ probability of the fit must be at least~0.02.
We define the difference between $\pi$ and the opening angle of the
two muons as the acollinearity~$\alpha$, and remove a
very small region of phase space where $\alpha$ is less than 5\unit{mrad}
to reduce contamination by
cosmic rays to a negligible level.

Photons are reconstructed using the particle flow~(PF) algorithm~\cite{PF2,PFJet}
that uses clustered energy deposits in ECAL.  The PF algorithm allows us to
reconstruct photons at relatively low~\ET and to maintain coherence
with the calculation of the isolation observables
described below. Photons that convert to electron-positron pairs are included in
this reconstruction.  Events selected for this analysis
must have at least one photon with $\ET > 5\GeV$, and the
separation of this photon with respect to the closest
muon must satisfy $0.05 < \DRmg \leq 3$.
Studies using the simulation show that photons with $\DRmg < 0.05$
are difficult to reconstruct reliably
due to the energy deposition left by the muon, and no signal
photons with $\DRmg > 3$ are expected. If an event has
more than one photon satisfying this selection criteria,
we select the one with the highest~\ET.  In events in which
one photon is selected, a second photon is present 15\%
of the time; however, these extra photons are expected to be mostly background,
since the fraction of events with a second FSR photon in simulation is approximately 0.5\%.
More details about these background photons are given in
Section~\ref{sec:backgrounds}.

\begin{table}[htb]
\centering
\topcaption{\label{tab:cuts}
Summary of kinematic and fiducial event requirements
}
\begin{scotch}{ll}
Object & Requirement \cr
\hline
Leading muon   & $\PT > 31\GeV$ and $\abs{\eta}< 2.4$ \cr
Trailing muon  & $\PT > 9\GeV$ and $\abs{\eta}< 2.4$ \cr
Acollinearity  & $\alpha > 0.005$ radians \cr
Photon         & $\ET > 5\GeV$, \cr
               & $\abs{\eta}< 2.4$ but not $1.4 < \abs{\eta}< 1.6$;\cr
               & $0.05 < \DRmg \leq 3$  \cr
Signal region  & $30 < \Mmm < 87\GeV$ \cr
Control region & $89 < \Mmm < 100\GeV$ \cr
\end{scotch}

\end{table}

All three particles emitted in the \Z~boson decay---the
two muons and the photon---are usually isolated
from other particles produced in the same bunch crossing.
We can reduce backgrounds substantially by imposing
appropriate isolation requirements.  The isolation
quantities, $\Imu$ for the muons and $\Igamma$ for the photon,
are based on the scalar \PT sums of reconstructed PF particles
within a cone around the muon or photon direction.  The cone size for both
muons and photons is $\Delta R = 0.3$.  The muon \PT is not included in the sum for $\Imu$,
and the photon \ET is not included in the sum for $\Igamma$;
these isolation quantities are meant to represent the energy carried by
particles originating from the main primary vertex close to the given muon or photon.  For a well-isolated
muon or photon, $\Imu$ or $\Igamma$
should be small.

Special care is
taken to avoid inefficiencies and biases occurring
when the FSR photon falls close to the muon; in such
cases the muon and the photon may appear, superficially,
to be nonisolated.  To avoid this effect, we exclude any PF
photon from the muon isolation sum.  Furthermore, since the photon can
convert and produce
charged particle tracks that cannot always be unambiguously
identified as an $\Pep\Pem$
pair, we exclude from the muon isolation sum charged tracks
that lack hits in the pixel detector or that have
$\PT < 0.5\GeV$.  Finally, any particle that lands
in a cone of $\Delta R < 0.2$ around a PF photon is
excluded from the muon isolation sum.   After these modifications to
the muon isolation variable, the efficiency of
the isolation requirement is flat (98\%) for all $\DRmg$ and is
higher than the efficiency of the unmodified isolation
requirement by about~0.5\%.  Adding these modifications does not significantly
increase the backgrounds.

The instantaneous luminosity of the LHC was sufficiently high
that each bunch crossing resulted in multiple  $\Pp\Pp$~collisions
(8.2~on average).
The extraneous pp~collisions are referred to as ``pileup''
and must be taken into consideration when defining and
calculating the muon and photon isolation variables.
Charged hadrons, electrons, and muons coming from pileup
can be identified by checking their point of origin along
the beam line, which will typically not coincide with the primary vertex
from which the muons originate.  When summing the contributions
of charged hadrons, electrons, and muons to the isolation
variable, those coming from pileup are excluded.  This
distinction is not possible for photons and neutral
hadrons, however.  Instead, an estimate $I_{\mathrm{p}}$ of the contribution
of photons and neutral hadrons to the sum is made: we use
one-half of the (already excluded) contribution from
charged hadrons within the isolation cone.  This estimate is subtracted from the
sum of contributions from photons and neutral hadrons;
if the result is negative, we then use a net contribution
of zero.

We designate by $I_{\mathrm{h}^\pm}$ the sum of \PT for charged particles
that are not excluded from the isolation sum.  We let $I_{\mathrm{em}}$ and
$I_{\mathrm{h^0}}$ stand for the
sums over the \PT of all photons and neutral hadrons, and
$I_{\mathrm{p}}$ for the estimate of the pileup
contribution to $I_{\mathrm{em}}$ and $I_{\mathrm{h^0}}$.
The muon isolation variable is, then,
\begin{equation}
 \Imu = \left( I_{\mathrm{h}^\pm} + I_{\mathrm{h^0}} \right) / \PT .
\end{equation}
Note that the sum is normalized to the \PT of the muon.
We require $\Imu < 0.2$ for both muons.

The photon isolation variable is calculated as above,
except that the muons are not included in the sum,
and there is no special exclusion of charged tracks
near the photon:
\begin{equation}
 \Igamma = I_{\mathrm{h}^\pm} +
   \max(I_{\mathrm{em}} + I_{\mathrm{h^0}}
       - I_{\mathrm{p}} , 0 )  .
\end{equation}
We require $\Igamma < 6\GeV$.

The emission of FSR photons in \Z~boson decays reduces the
momenta of the muons.  Consequently, the dimuon mass $\Mmm$
tends to be lower than $M_Z$, the nominal mass of the \Z~boson.
Simulations indicate that, for most of the signal, $\Mmm < 87\GeV$,
due to the requirement $\ET > 5$\GeV for the photon.
They also show that the $\Mmm$ distribution for radiative
decays $\Z \to \mpmm\gamma$ ends at $\Mmm \approx 30\GeV$.
A requirement $\Mmm > 30\GeV$ also helps to avoid a kinematic region
in which the acceptance is difficult to model.  Therefore, our signal region
is defined by $30 < \Mmm < 87\GeV$.  We also
define a control region by $89 < \Mmm < 100\GeV$, where
the contribution of FSR photons is quite small (below~0.5\%).
The numbers of events we select are \NSIGNAL in the signal region
and \NCONTROL in the control region.

\section{Background estimation}
\label{sec:backgrounds}

Nearly all selected events have two prompt muons from the DY process.
Backgrounds come mainly from ``nonprompt'' photons, which may be genuine
photons produced in the decays of light mesons (such as $\Pgpz$
and $\eta$), a pair of overlapping photons that cannot be distinguished
from a single photon, and photons from pileup.  We study these
backgrounds with simulated DY events and apply corrections so that
the simulation reproduces the data distributions, as described
in detail below.

Some events come from processes other than DY, such as $\ttbar$ and diboson production.
These backgrounds are very small and
are estimated using the simulation. Similarly, a small background
from the DY production of $\Pgt^+\Pgt^-$ is also estimated from simulation.
The background from multijet events, including events with
a $\Wpm \to \mu^\pm\nu$ decay, is estimated using events
with same-sign muons.  Backgrounds from simultaneous nonprompt muon and nonprompt
photon sources are negligible.  The composition of the signal sample
is given in Table~\ref{tab:composition}.

\begin{table}[htb]
\centering
\topcaption{\label{tab:composition}
Composition of the signal sample.
The simulation has been tuned to reproduce
the data in the control region.}
\begin{scotch}{lr}
Process & Fraction \cr
\hline
Signal         &  77.1\% \cr
DY with a nonprompt photon & 9.5\% \cr
Pileup        & 11.2\% \cr
$\ttbar$       &  0.6\% \cr
$\tau^+\tau^-$ &  0.3\% \cr
Dibosons      &  0.2\% \cr
Multijets      &  1.1\% \cr
\end{scotch}

\end{table}

The control region is dominated by nonprompt photons whose
kinematic distributions ($\ET$, $\eta$, $\DRmg$) are
nearly identical to nonprompt photons in the signal region.
Quantitative comparisons of data and simulation
revealed significant discrepancies in the control region
that prompted corrections to the simulation, which we now explain.

The \POWHEG{}+\PYTHIA sample does not reproduce the number of jets
per event well, so we apply weights to the simulated events
as a function of the number of reconstructed jets in each event.
For this purpose, jets are reconstructed from
PF objects using the anti-\kt algorithm~\cite{ANTIKT} with a
size parameter $R = 0.5$.  We consider jets with
$\PT > 20\GeV$ and $\abs{\eta}< 2.4$ that
do not overlap with the muons or the photon.

Studies of $\Igamma$ for events
in the control region reveal small discrepancies
in the multiplicity and \PT spectra of charged hadrons
included in the sum.  We apply weights to the simulated
events to bring the multiplicities into agreement, and
we impose $\PT > 0.5\GeV$ on charged hadrons.  The
simulated $\Igamma$ distributions match those in
data very well after applying these weights.

Finally, the $\ET$ and $\eta$ distributions of nonprompt photons
in the simulation deviate from those in the data.  We fit
simple analytic functions to the ratios of data to simulated \ET and $\eta$
distributions and define
a weight as the product of those functions.  We check
that this factorization is valid (\ie, that the \ET
correction is the same for different narrow ranges of $\eta$,
and vice versa).

\begin{figure}[htb]
\centering
\includegraphics[width=0.49\textwidth]{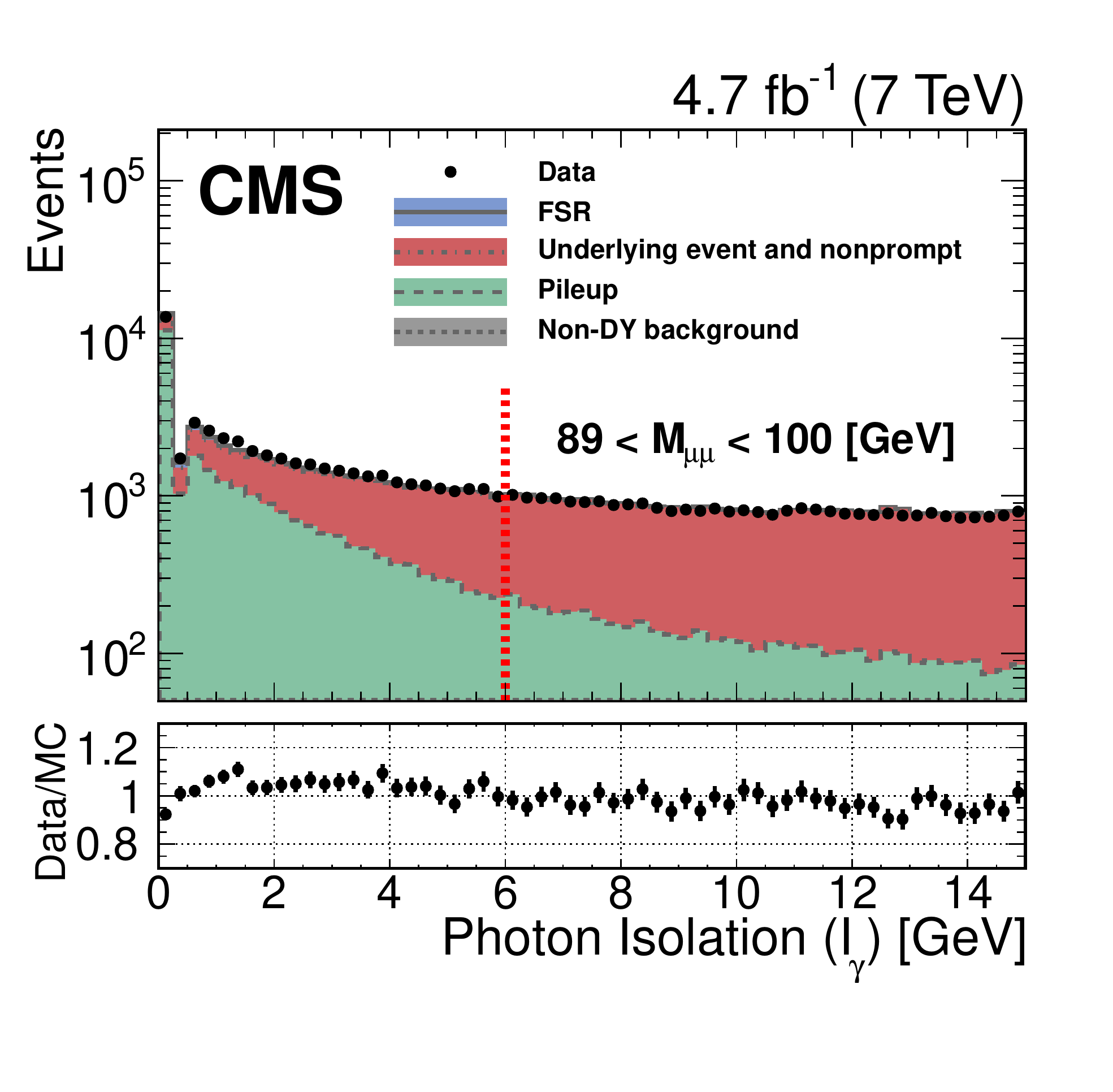}
\includegraphics[width=0.49\textwidth]{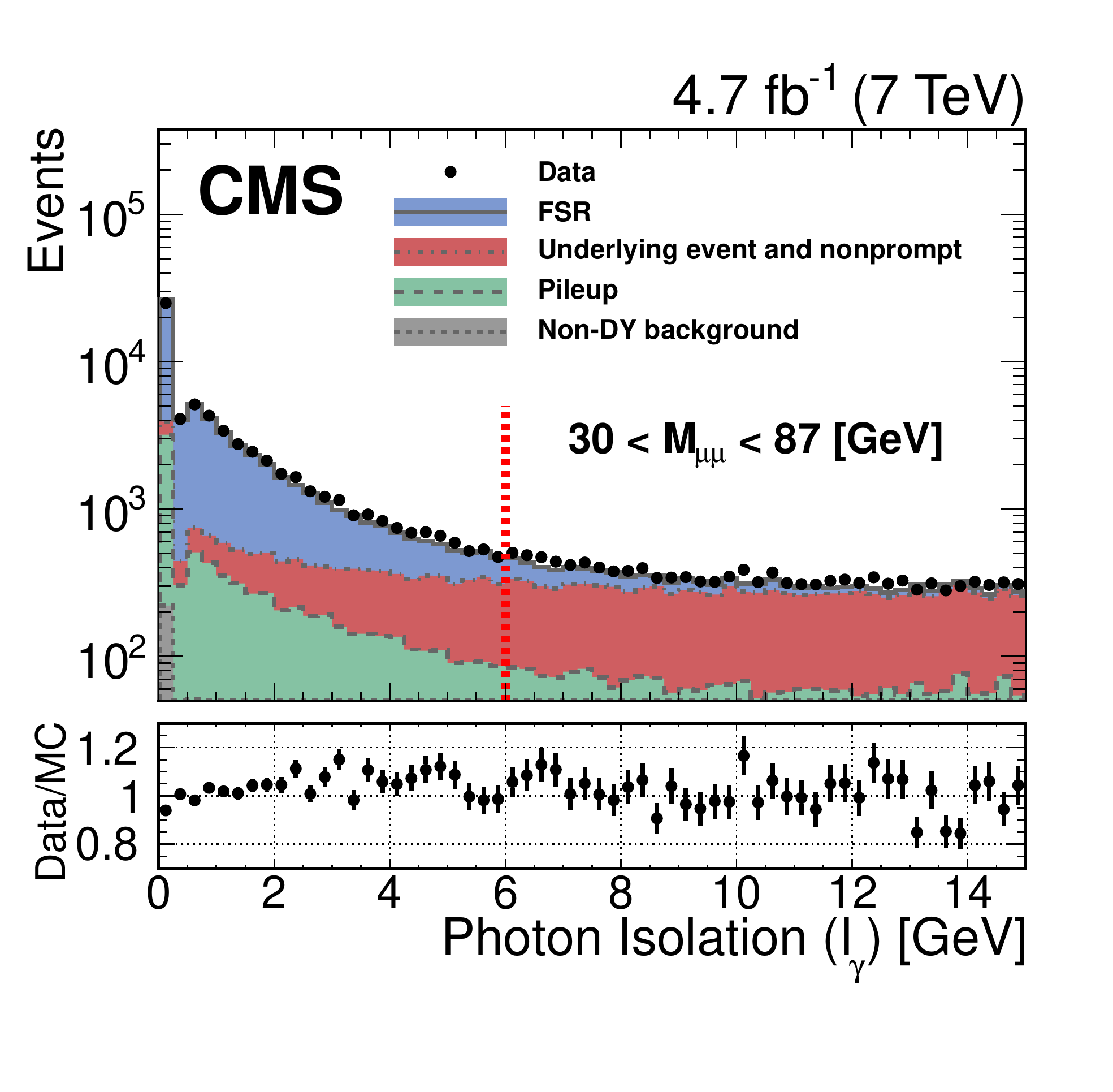}
\caption[.]{\label{fig:ISO}
Distributions of the photon isolation variable $\Igamma$
for the control region~(\cmsLeft) and for the signal region~(\cmsRight)
after all corrections have been applied.  The bottom panels display
the ratio of data to the MC expectation.
The requirement for FSR photons is $\Igamma < 6\GeV$.}
\end{figure}

After these three corrections (for jet multiplicity,
for the spectrum of charged hadrons in the isolation sums,
and for the \ET and $\eta$ of the nonprompt photon),
the simulation matches the data in all kinematic
distributions in the control region, an example of which is shown in
Fig.~\ref{fig:ISO},~\cmsLeft.  The total change in the background estimate due to these
corrections is approximately 5 to 10\%. We use the simulation
with these weights to model the small background in the signal
region (Fig.~\ref{fig:ISO},~\cmsRight).

Given the definition of the signal region, the contribution
of photons emitted in the initial state is very small
(on the order of $4 \times 10^{-4}$ as determined from the \POWHEG{}+\PYTHIA sample)
and is counted as signal.

\section{Correcting for detector effects}
\label{sec:unfold}

Our goal is to measure differential cross sections in a form that
is optimal for testing FSR calculations.  Therefore we are
obliged to remove the effects of detector resolution
and efficiency (including reconstruction, isolation, and trigger efficiency).
The corrections for the muons follow the techniques
developed for the DY cross section measurements~\cite{DY}.
The corrections for photons are applied using an unfolding
technique, as discussed in this section.

We apply small corrections to the muon momentum scale as a function
of muon \PT, $\eta_\mu$, and $\phi_\mu$~\cite{ROCHESTER};
they have almost no impact on our measurements.  The muon reconstruction efficiency,
(taken from simulation and corrected to match the data as a function of \PT and $\eta_\mu$)
is taken into account by applying weights on a per-event basis.
We do not correct for the approximately 0.5\%
increase in the isolation efficiency coming from the way we handle
FSR photons falling within the muon isolation cone.

The energy scale and efficiencies for photons are more central
to our task.  Most PF photon energies correspond to the true
energies within a few percent.  However, in about 13\%
of the cases the photon energy is significantly underestimated.
The simulation reproduces this effect
very well.  We construct a ``response'' matrix that relates the
PF energy to the true energy as a function of $\eta_\gamma$ and $\DRmg$.
The angular quantities $\eta_\gamma$ and $\DRmg$ are themselves
well measured.  We use the iterative D'Agostini method of unfolding~\cite{BayesUnfolding}
as implemented in the \textsc{RooUnfold}
package~\cite{RooUnfold}.   By default, we unfold in the three
quantities $\ET$, $\eta_\gamma$, and $\DRmg$ simultaneously after subtracting
backgrounds; as a
cross check we also unfold the \ET and $\DRmg$ distributions
one at a time, and we also use a single-value decomposition method~\cite{SVD}.
All results are consistent with each other.  To verify the independence of the unfolded
result on the assumed spectra, we distort the FSR model
in an arbitrary manner when reconstructing the response matrix.
The unfolded result is no different than the original one
we obtained.  A closure test in which the simulation is treated as data and undergoes
the same unfolding procedure indicates no deviation greater
than 1.5\%.

The unfolding procedure corrects for the photon reconstruction
and isolation efficiency.   It also corrects for bias in
the PF photon energy assuming that such a bias is reproduced in the simulation.
Verification of the
photon efficiencies and
energy scale in data with respect to the simulation are discussed in Section~\ref{sec:systematics}.

\section{Systematic uncertainties}
\label{sec:systematics}

Systematic uncertainties are assigned to each step of the analysis procedure
using methods detailed in this section.
Tables~\ref{tab:syst_dxdet} and~\ref{tab:syst_dxddr} present
a summary of these uncertainties, which are similar in
magnitude to, or somewhat larger than the statistical
uncertainties, depending on the photon~$\ET$.

\begin{table*}[htbp]
\centering
\caption{\label{tab:syst_dxdet}
Relative systematic uncertainties for $\DXDET$ (in percent).}
\cmsTable{\linewidth}{%
 \begin{scotch}{lD{.}{.}{2.1}cccccc|c}
 Kinematic   &  \multicolumn{1}{c}{Background}  & Muon       & Photon    & Photon \ET & Photon     &  Pileup &  Unfolding & Total \\
 requirement &  \multicolumn{1}{c}{estimation}  & efficiency & \ET-scale & resolution &  efficiency &  photons  &     &  \\
 {[}GeV{]} & & & & & & & & \\
\hline
\multicolumn{9}{c}{$0.05 < \DRmg \leq 3$} \\
\hline
    $5\phantom{0} < \ET \leq 10$ & 2.7  & 3.0  & 0.5 & 1.0 & 2.0 & \hphantom{$<$}1.5    & 1.4 & 5.1 \\
          $10 < \ET \leq 15$   & 1.3 & 2.5  & 0.5 & 0.5 & 1.0 & \hphantom{$<$}0.4    & 1.4 & 3.4 \\
          $15 < \ET \leq 20$   & 0.9 & 2.5  & 0.5 & 0.5 & 1.3 & \hphantom{$<$}0.1    & 1.4 & 3.3 \\
          $20 < \ET \leq 25$   & 0.8 & 2.7  & 0.5 & 0.5 & 1.4 & $<$0.1 & 1.4 & 3.5 \\
          $25 < \ET \leq 30$   & 0.7 & 3.3  & 0.5 & 0.5 & 1.5 & $<$0.1 & 1.4 & 4.0 \\
          $30 < \ET \leq 40$   & 1.0 & 4.3  & 0.5 & 0.5 & 1.1 & \hphantom{$<$}0.1    & 1.4 & 4.8 \\
          $40 < \ET \leq 50$   & 2.9 & 4.4  & 1.0 & 0.5 & 2.8 & \hphantom{$<$}0.5    & 1.4 & 6.3 \\
          $50 < \ET \leq 75$   & 7.2 & 4.5  & 1.0 & 0.5 & 2.0 & \hphantom{$<$}0.6    & 1.4 & 8.9 \\
          $75 < \ET \leq 100$  & 15.3 & 4.5 & 3.0 & 1.0 & 6.9 & \hphantom{$<$}1.1    & 1.4 & 17.8 \\
 \hline
\multicolumn{9}{c}{$0.05 < \DRmg \leq 0.5$} \\
\hline
 $5\phantom{0} < \ET \leq 10$ & 0.8  & 2.1  & 0.5 & 1.0 & 2.0 & \hphantom{$<$}0.1    & 1.4 & 3.5 \\
         $10 < \ET \leq 15$   & 0.4 & 2.0  & 0.5 & 0.5 & 1.0  & $<$0.1 & 1.4 & 2.8 \\
         $15 < \ET \leq 20$   & 0.3 & 2.2  & 0.5 & 0.5 & 1.3  & $<$0.1 & 1.4 & 3.1 \\
         $20 < \ET \leq 25$   & 0.3 & 2.5  & 0.5 & 0.5 & 1.4  & $<$0.1 & 1.4 & 3.3 \\
         $25 < \ET \leq 30$   & 0.2 & 3.2  & 0.5 & 0.5 & 1.5  & $<$0.1 & 1.4 & 3.9 \\
         $30 < \ET \leq 40$   & 0.3 & 4.3  & 0.5 & 0.5 & 1.1  & $<$0.1 & 1.4 & 4.7 \\
         $40 < \ET \leq 50$   & 0.9 & 3.9  & 1.0 & 0.5 & 2.8  & $<$0.1 & 1.4 & 5.2 \\
         $50 < \ET \leq 75$   & 2.3 & 3.0  & 1.0 & 0.5 & 2.0  & $<$0.1 & 1.4 & 4.6 \\
         $75 < \ET \leq 100$  & 4.9 & 3.1  & 3.0 & 1.0 & 6.9  & \hphantom{$<$}0.8    & 1.4 & 9.7 \\
 \hline
\multicolumn{9}{c}{$0.5 < \DRmg \leq 3$} \\
\hline
 $5\phantom{0} < \ET \leq 10$ & 6.4  & 4.7  & 0.5 & 1.0 & 2.0 & \hphantom{$<$}3.8   & 1.4 & 9.2 \\
         $10 < \ET \leq 15$   & 2.8 & 3.2  & 0.5 & 0.5 & 1.0  & \hphantom{$<$}0.8   & 1.4 & 4.7 \\
         $15 < \ET \leq 20$   & 1.9 & 2.8  & 0.5 & 0.5 & 1.3  & \hphantom{$<$}0.3   & 1.4 & 4.0 \\
         $20 < \ET \leq 25$   & 1.7 & 3.0  & 0.5 & 0.5 & 1.4  & $<$0.1 & 1.4 & 4.0 \\
         $25 < \ET \leq 30$   & 1.6 & 3.4  & 0.5 & 0.5 & 1.5  & \hphantom{$<$}0.1   & 1.4 & 4.3 \\
         $30 < \ET \leq 40$   & 2.3 & 4.4  & 0.5 & 0.5 & 1.1  & \hphantom{$<$}0.2   & 1.4 & 5.3 \\
         $40 < \ET \leq 50$   & 6.5 & 5.1  & 1.0 & 0.5 & 2.8  & \hphantom{$<$}1.3   & 1.4 & 9.0 \\
         $50 < \ET \leq 75$   & 16.1 & 8.1  & 1.0 & 0.5 & 2.0 & \hphantom{$<$}2.0   & 1.4 & 18.4 \\
         $75 < \ET \leq 100$  & 34.5 & 6.2  & 3.0 & 1.0 & 6.9 & \hphantom{$<$}3.5   & 1.4 & 36.0 \\
 \hline
\multicolumn{9}{c}{$0.05 < \DRmg \leq 3$ and $\QT < 10\GeV$} \\
\hline
  $5\phantom{0} < \ET \leq 10$ & 1.4  & 2.2  & 0.5 & 1.0      & 2.0   & \hphantom{$<$}1.0 & 1.4 & 3.9 \\
         $10 < \ET \leq 15$   & 0.6 & 1.9  & 0.5 & 0.5 & 1.0  & \hphantom{$<$}0.1   & 1.4 & 2.8 \\
         $15 < \ET \leq 20$   & 0.4 & 2.1  & 0.5 & 0.5 & 1.3  & $<$0.1 & 1.4 & 3.0 \\
         $20 < \ET \leq 25$   & 0.4 & 2.4  & 0.5 & 0.5 & 1.4  & $<$0.1 & 1.4 & 3.3 \\
         $25 < \ET \leq 30$   & 0.5 & 3.5  & 0.5 & 0.5 & 1.5  & $<$0.1 & 1.4 & 4.1 \\
         $30 < \ET \leq 40$   & 0.6 & 5.1  & 0.5 & 0.5 & 1.1  & $<$0.1 & 1.4 & 5.5 \\
         $40 < \ET \leq 50$   & 7.3 & 4.7  & 1.0 & 0.5 & 2.8  & \hphantom{$<$}1.0   & 1.4 & 9.4 \\
         $50 < \ET \leq 75$   & 18.2 & 8.5  & 1.0 & 0.5 & 2.0 & \hphantom{$<$}4.4   & 1.4 & 20.8 \\
         $75 < \ET \leq 100$  & 38.9 & 6.4  & 3.0 & 1.0 & 6.9 & $<$0.1 & 1.4 & 40.2 \\
 \hline
\multicolumn{9}{c}{$0.05 < \DRmg \leq 3$ and $\QT > 50\GeV$} \\
\hline
 $5\phantom{0} < \ET \leq 10$ & 5.7  & 4.2  & 0.5 & 1.0 & 2.0 & \hphantom{$<$}1.8  & 1.4 & 7.8 \\
         $10 < \ET \leq 15$   & 3.0 & 3.3  & 0.5 & 0.5 & 1.0  & \hphantom{$<$}0.4  & 1.4 & 4.9 \\
         $15 < \ET \leq 20$   & 3.0 & 2.8  & 0.5 & 0.5 & 1.3  & \hphantom{$<$}0.3  & 1.4 & 4.6 \\
         $20 < \ET \leq 25$   & 2.3 & 2.7  & 0.5 & 0.5 & 1.4  & \hphantom{$<$}0.2  & 1.4 & 4.2 \\
         $25 < \ET \leq 30$   & 1.9 & 2.6  & 0.5 & 0.5 & 1.5  & \hphantom{$<$}0.1  & 1.4 & 3.9 \\
         $30 < \ET \leq 40$   & 2.9 & 2.9  & 0.5 & 0.5 & 1.1  & \hphantom{$<$}0.3  & 1.4 & 4.6 \\
         $40 < \ET \leq 50$   & 1.5 & 2.8  & 1.0 & 0.5 & 2.8  & \hphantom{$<$}0.2  & 1.4 & 4.6 \\
         $50 < \ET \leq 75$   & 3.9 & 2.8  & 1.0 & 0.5 & 2.0  & \hphantom{$<$}0.3  & 1.4 & 5.5 \\
         $75 < \ET \leq 100$  & 8.2 & 3.5  & 3.0 & 1.0 & 6.9  & \hphantom{$<$}0.2  & 1.4 & 11.8 \\
 \end{scotch}
 }

 \end{table*}

\begin{table*}[htbp]
\centering
\caption{\label{tab:syst_dxddr}
Relative systematic uncertainties for $\DXDDR$ (in percent).}
\cmsTable{\linewidth}{%
 \begin{scotch}{l*{7}{c}|c}
 Kinematic   &  Background  & Muon       & Photon    & Photon \ET & Photon     &  Pileup &  Unfolding & Total \\
 requirement &  estimation  & efficiency & \ET-scale & resolution &  efficiency &  photons &  & \\
\hline
\multicolumn{9}{c}{$\ET > 5.0\GeV$} \\
\hline
              $0.15 < \DRmg \leq 0.1$ & 0.7  & 2.4  & $<$0.1 & $<$0.1 & 1.0 & $<$0.1 & 1.4 & 3.0 \\
    $0.1\phantom{0} < \DRmg \leq 0.15$ & 0.6 & 2.3  & $<$0.1 & $<$0.1 & 1.1 & $<$0.1 & 1.4 & 3.0 \\
               $0.15 < \DRmg \leq 0.3$ & 0.4 & 2.3  & $<$0.1 & $<$0.1 & 1.0 & $<$0.1 & 1.4 & 2.9 \\
     $0.3\phantom{0} < \DRmg \leq 0.5$ & 0.5 & 2.3  & $<$0.1 & $<$0.1 & 1.0 &\hphantom{$<$}0.1   & 1.4 & 3.0 \\
     $0.5\phantom{0} < \DRmg \leq 0.8$ & 1.1 & 2.6  & $<$0.1 & $<$0.1 & 1.0 &\hphantom{$<$}0.6   & 1.4 & 3.4 \\
     $0.8\phantom{0} < \DRmg \leq 1.2$ & 2.2 & 3.2  & $<$0.1 & $<$0.1 & 1.1 &\hphantom{$<$}1.1   & 1.4 & 4.4 \\
     $1.2\phantom{0} < \DRmg \leq 1.6$ & 4.1 & 3.7  & $<$0.1 & $<$0.1 & 1.1 &\hphantom{$<$}1.7   & 1.4 & 6.1 \\
     $1.6\phantom{0} < \DRmg \leq 2.0$ & 6.6 & 4.9  & $<$0.1 & $<$0.1 & 1.1 &\hphantom{$<$}2.8   & 1.4 & 8.8 \\
     $2.0\phantom{0} < \DRmg \leq 3.0$ & 18.3 & 9.9  & $<$0.1 & $<$0.1 & 1.3&\hphantom{$<$}7.9   & 1.4 & 22.3 \\
\hline
\multicolumn{9}{c}{$\ET > 5.0\GeV$ and $\QT < 10\GeV$} \\
\hline
              $0.15 < \DRmg \leq 0.1$ & 0.2  & 2.1  & $<$0.1 & $<$0.1 & 1.0 & $<$0.1 & 1.4 & 2.8 \\
    $0.1\phantom{0} < \DRmg \leq 0.15$ & 0.2 & 2.2  & $<$0.1 & $<$0.1 & 1.1 & $<$0.1 & 1.4 & 2.8 \\
               $0.15 < \DRmg \leq 0.3$ & 0.1 & 2.1  & $<$0.1 & $<$0.1 & 1.0 & $<$0.1 & 1.4 & 2.7 \\
     $0.3\phantom{0} < \DRmg \leq 0.5$ & 0.3 & 2.2  & $<$0.1 & $<$0.1 & 1.0 &\hphantom{$<$}0.1   & 1.4 & 2.8 \\
     $0.5\phantom{0} < \DRmg \leq 0.8$ & 0.7 & 2.4  & $<$0.1 & $<$0.1 & 1.0 &\hphantom{$<$}0.3& 1.4 & 3.0 \\
     $0.8\phantom{0} < \DRmg \leq 1.2$ & 1.3 & 2.5  & $<$0.1 & $<$0.1 & 1.1 &\hphantom{$<$}0.6& 1.4 & 3.4 \\
     $1.2\phantom{0} < \DRmg \leq 1.6$ & 2.2 & 2.7  & $<$0.1 & $<$0.1 & 1.1 &\hphantom{$<$}1.0& 1.4 & 4.1 \\
     $1.6\phantom{0} < \DRmg \leq 2.0$ & 3.8 & 3.1  & $<$0.1 & $<$0.1 & 1.1 &\hphantom{$<$}2.1& 1.4 & 5.6 \\
     $2.0\phantom{0} < \DRmg \leq 3.0$ & 15.9 & 7.4  & $<$0.1 & $<$0.1 & 1.3&\hphantom{$<$}9.0 & 1.4 & 19.8 \\
\hline
\multicolumn{9}{c}{$\ET > 5.0\GeV$ and $\QT > 50\GeV$} \\
\hline
              $0.15 < \DRmg \leq 0.1$ & 1.8  & 2.5  & $<$0.1 & $<$0.1 & 1.0 & $<$0.1 & 1.4 & 3.6 \\
    $0.1\phantom{0} < \DRmg \leq 0.15$ & 1.1 & 2.3  & $<$0.1 & $<$0.1 & 1.1 & $<$0.1 & 1.4 & 3.1 \\
               $0.15 < \DRmg \leq 0.3$ & 1.5 & 2.4  & $<$0.1 & $<$0.1 & 1.0 & $<$0.1 & 1.4 & 3.4 \\
     $0.3\phantom{0} < \DRmg \leq 0.5$ & 1.7 & 2.4  & $<$0.1 & $<$0.1 & 1.0 &\hphantom{$<$}0.1 & 1.4 & 3.4 \\
     $0.5\phantom{0} < \DRmg \leq 0.8$ & 2.6 & 2.9  & $<$0.1 & $<$0.1 & 1.0 &\hphantom{$<$}0.7 & 1.4 & 4.4 \\
     $0.8\phantom{0} < \DRmg \leq 1.2$ & 4.2 & 3.8  & $<$0.1 & $<$0.1 & 1.1 &\hphantom{$<$}1.4 & 1.4 & 6.1 \\
     $1.2\phantom{0} < \DRmg \leq 1.6$ & 9.1 & 5.2  & $<$0.1 & $<$0.1 & 1.1 &\hphantom{$<$}1.9 & 1.4 & 10.8 \\
    $1.6\phantom{0} < \DRmg \leq 2.0$ & 14.9 & 7.4  & $<$0.1 & $<$0.1 & 1.1 &\hphantom{$<$}3.4 & 1.4 & 17.1 \\
   $2.0\phantom{0} < \DRmg \leq 3.0$ & 22.3 & 10.3  & $<$0.1 & $<$0.1 & 1.3 &\hphantom{$<$}5.1 & 1.4 & 25.1 \\
 \end{scotch}
 }

 \end{table*}

The muon efficiency, taken from simulation, is corrected as a function
of \PT and $\eta$ using a method derived from data and described in Ref.~\cite{DY}.
The statistical uncertainties for these corrections constitute
a systematic uncertainty, which we also take from Ref.~\cite{DY}.
In addition, we assign a 0.5\% uncertainty to account for
the modifications of the standard isolation variable.  We propagate
the uncertainty in the muon efficiency by shifting the per-event weights
up and down by one unit of systematic uncertainty.

\begin{figure}[htb]
\centering
\includegraphics[width=0.49\textwidth]{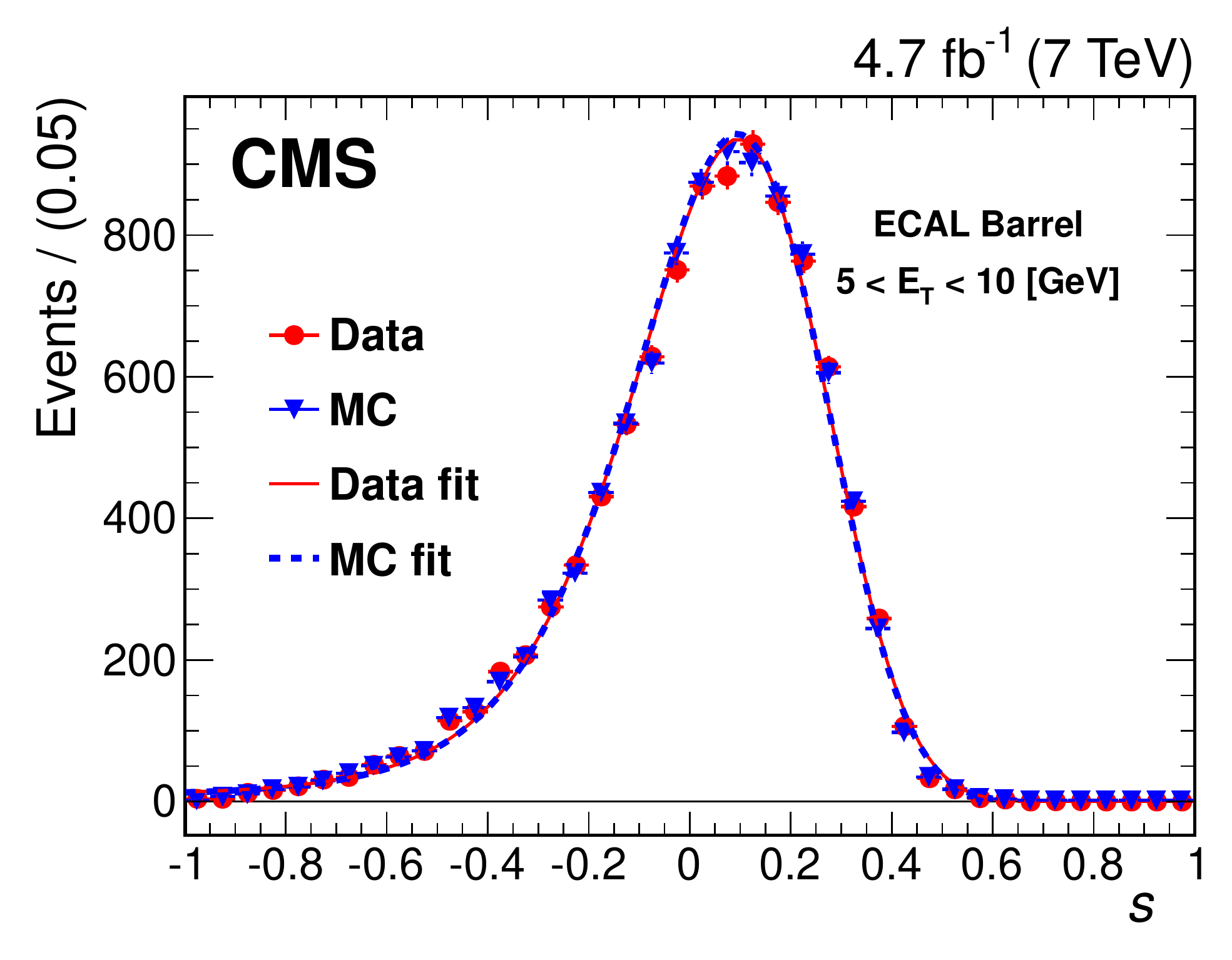}
\includegraphics[width=0.49\textwidth]{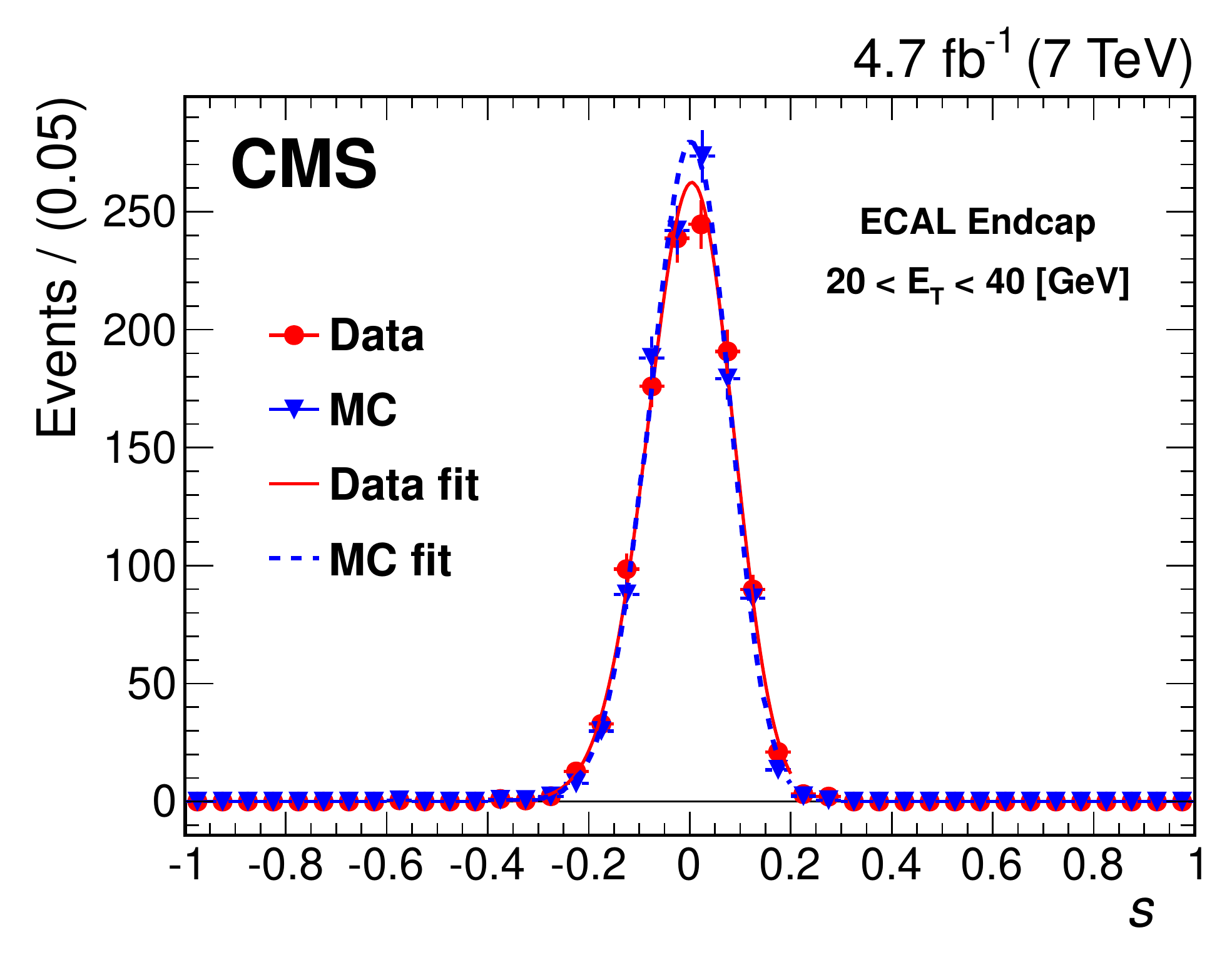}
\caption[.]{\label{fig:s}
Two examples of an $s$ distribution $s = 1 - (M^2_{\mu\mu\gamma} - M^2_{\mu\mu})/(M^2_Z - M^2_{\mu\mu})$
fit with a skewed Gaussian as described in the text.
The \cmsLeft and \cmsRight plots pertain to photons in the ECAL barrel with
$5 < \ET < 10\GeV$ and in the ECAL endcaps with $20 < \ET < 40\GeV$,
respectively. The circle points and solid curve represent
the data and the triangle points and dotted curve represent the simulation.}
\end{figure}

The photon $\ET$ scale is potentially an important source of
systematic uncertainty although simulations indicate that the bias in PF photon energy is negligibly
small.  We verify
the fidelity of the simulation by introducing an extra requirement,
$0.05 < \DRmg \leq 0.9$, which gives us a high-purity subset of signal
events in which the energy of the photon can be estimated from
just the muon kinematics.  We refer to this estimate as $\EgammaKIN$.  The quantity
$s = 1 - (M^2_{\mu\mu\gamma} - M^2_{\mu\mu})/(M^2_Z - M^2_{\mu\mu})
\approx 1 - \EgammaPF / \EgammaKIN$
is distributed as a skewed Gaussian with a mean close to zero.
We conducted detailed quantitative studies of the $s$ distribution
in bins of $\ETgammaPF$, separately in the barrel and
endcaps.  We fit the distributions to a Gaussian-like function
in which the width parameter is itself a function of~$s$,
namely, $\sigma(s) = c ( 1 + \re^{bs} )$, with $b$ and $c$ as free parameters.
Examples are given in Fig.~\ref{fig:s}.
Overall, the simulation reproduces the $s$ distributions in data very well.
We derive some small corrections from the differences in data and simulation as a function of $\ETgammaPF$
and construct an alternate response matrix.
The unfolded spectrum we obtained with this alternate response
matrix differs from the original by less than 0.2\% for
$\ET < 40\GeV$, by less than 1\% for $\ET < 75\GeV$,
and by less than 3\% in the highest \ET bin.   We assign respective
systematic uncertainties of 0.5\%, 1\%, and 3\% for
these three \ET ranges to account for the photon energy scale uncertainty.

The photon energy resolution uncertainty is well constrained by studies with
electrons and FSR photons~\cite{EGAMMA}.
To assess the impact of the uncertainty in the resolution,
we degrade the photon energy resolution in simulated events
by adding in quadrature a 1\% term to the nominal resolution
and construct a new response matrix.
The differences in the unfolded spectrum relative to the
default response matrix are small, and we take these differences
as the systematic uncertainty due to photon energy resolution.

Efficiency corrections for photons are applied as part of the unfolding
procedure described in Section~\ref{sec:unfold} and are derived from the simulation.
We verify these corrections using the data in the following way.
An isolated FSR photon with $\ET > 5\GeV$ nearly always produces
a cluster in the ECAL.
We define an efficiency to reconstruct and select PF photons given such
isolated clusters.
This efficiency rises from 60\% for $\ET$ between 5--10\GeV to
approximately 90\% for $\ET > 50\GeV$ and is nearly the same in data and simulation.
We take the difference,
added in quadrature to the statistical uncertainties of
the efficiencies, as the systematic uncertainty.

As described briefly in Section~\ref{sec:unfold}, the unfolding procedure has been
cross-checked in several ways.
To assess a systematic uncertainty due to unfolding, we use the
small discrepancies observed in the closure test.

The uncertainty in the background estimate is dominated by the
uncertainties associated with the corrections that we obtained
from the control region (Section~\ref{sec:backgrounds}).
The statistical uncertainty in the weights for jet multiplicity
has a negligible impact, as does the correction for
charged hadrons in the photon isolation cone.
The parameterized functions to correct the photon distributions
in \ET and $\eta$ carry statistical uncertainties that we propagate
to the measured cross sections through simplified MC models.
Since the nonprompt photon \ET, $\eta$, and $\DRmg$ distributions in
the control and signal regions are indistinguishable, we do not
assess any uncertainty in the modeling.

The uncertainties in the non-DY backgrounds ($\ttbar$ and diboson
production) are obtained from the uncertainties in the theoretical
cross sections, the luminosity, and the statistical uncertainty
in the simulated event samples.  We assign 50\% uncertainty
to the W$ + $jets and multijet background estimates, which
are quite small.

The systematic uncertainty from the simulation of pileup
depends primarily on the assumed
cross section for additional pp~collisions (roughly the same
as the minimum-bias cross section)~\cite{pileup}.  We vary the value of
this cross section by 5\% and evaluate the impact
on the unfolded spectra.

The uncertainty in the integrated luminosity is 2.2\%~\cite{LUMI}.

Theoretical uncertainties have been calculated and pertain
to the reported theoretical prediction only.  We propagated the
uncertainty due to parton distribution functions (PDFs)
using the prescription of Ref.~\cite{Whalley:2005nh}.
We vary the factorization/renormalization scale parameters by a factor of 2
to estimate associated scale uncertainties introduced due to neglected higher-order quantum
corrections.  Finally, we include the MC
statistical uncertainty.

\section{Results}
\label{sec:results}

The differential cross sections are obtained by subtracting the
estimated backgrounds from the observed distributions, unfolding
the result, and dividing by the bin width and the integrated
luminosity, $\mathcal{L} = \THELUMI$.  No acceptance correction
is applied, so these cross sections are defined relative
to the kinematic and fiducial requirements listed in Table~\ref{tab:cuts}.

\begin{figure}[htb]
\centering
\includegraphics[width=0.49\textwidth]{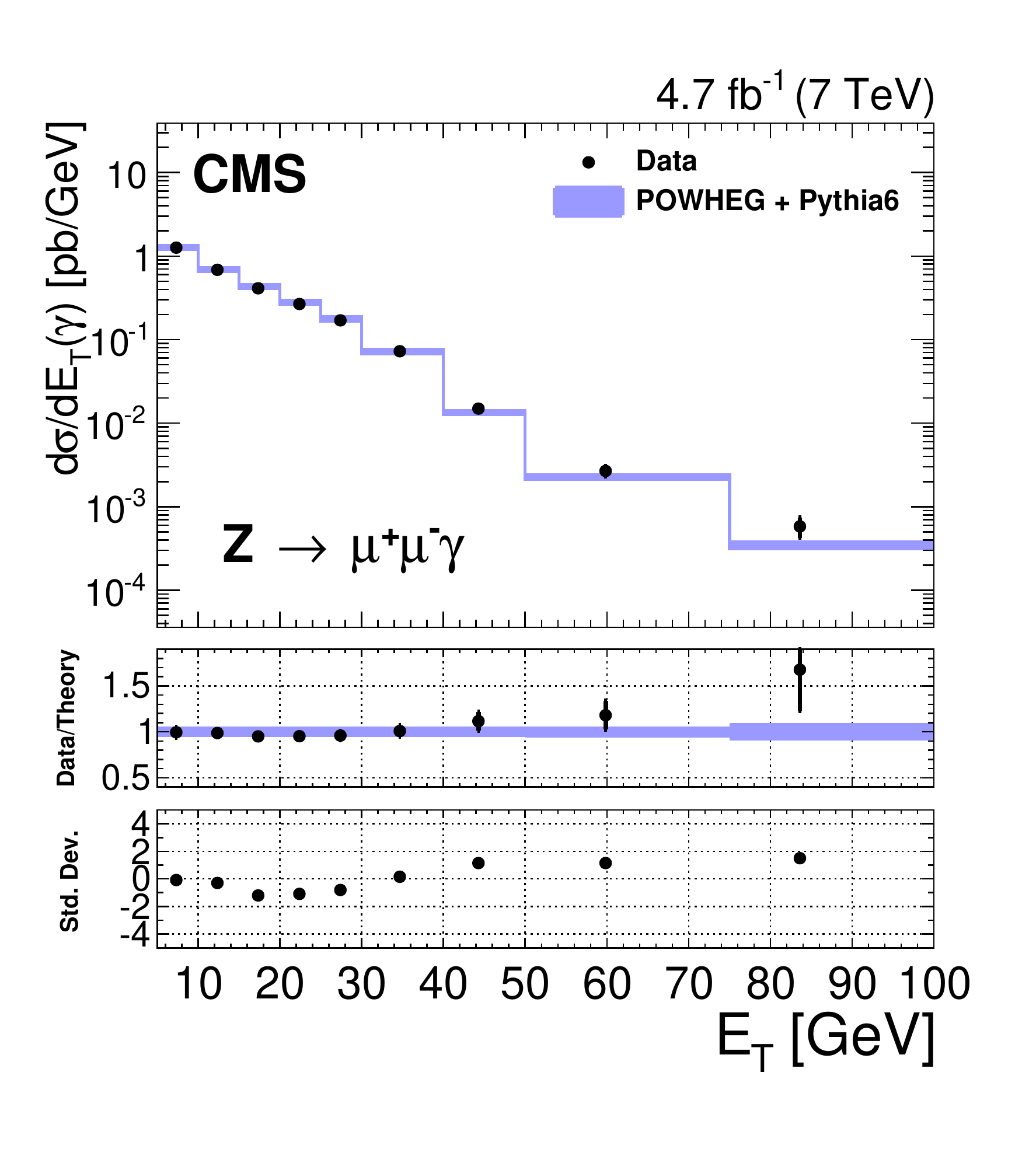}
\includegraphics[width=0.49\textwidth]{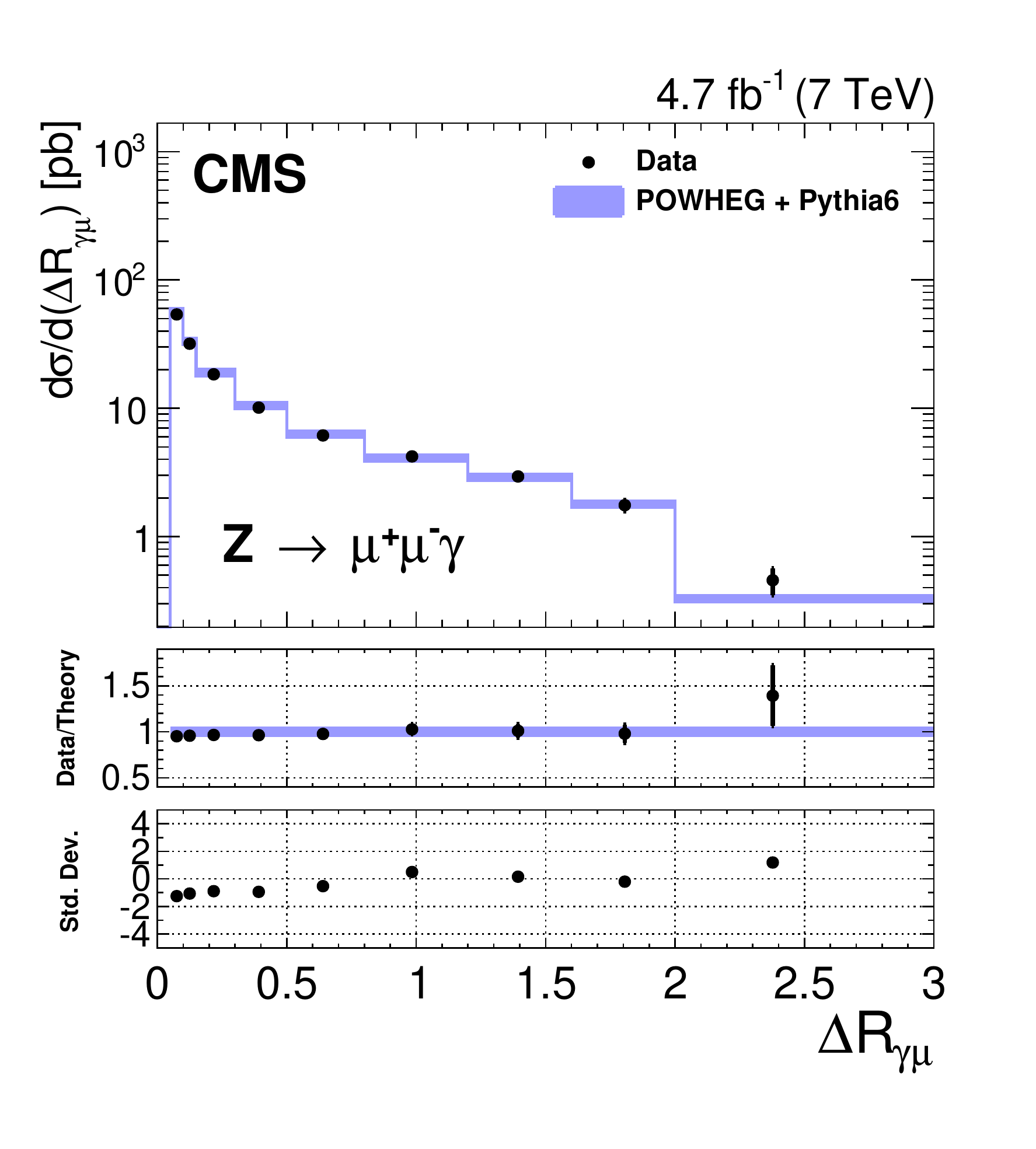}
\caption[.]{\label{fig:results1}
Measured differential cross sections $\DXDET$~(\cmsLeft)
and $\DXDDR$~(\cmsRight).
In the upper panels, the dots with error bars represent the data, and the shaded
bands represent the \POWHEG{}+\PYTHIA calculation including theoretical
uncertainties.  
The central panels display the ratio of data to the MC expectation.
The lower panels show the standard deviations of the measurements
with respect to the calculation.
A bin-centering procedure has been applied.}
\end{figure}

The measured differential cross sections $\DXDET$ and $\DXDDR$
are displayed in Fig.~\ref{fig:results1} and listed in
Tables~\ref{tab:dxdet} and~\ref{tab:dxddr}.
A bin-centering correction is applied
following the method of Ref.~\cite{BINCENTERING};
the abscissa of each point is based on the integral of
the simulation across the bin and on the bin width.
The shaded region represents the prediction and uncertainty from \POWHEG{}+\PYTHIA,
obtained at the parton level: only the requirements in
Table~\ref{tab:cuts} have been applied to the generator-level
muons and photons.  The agreement with data is good.

\begin{figure*}[htb]
\centering
\includegraphics[width=0.49\textwidth]{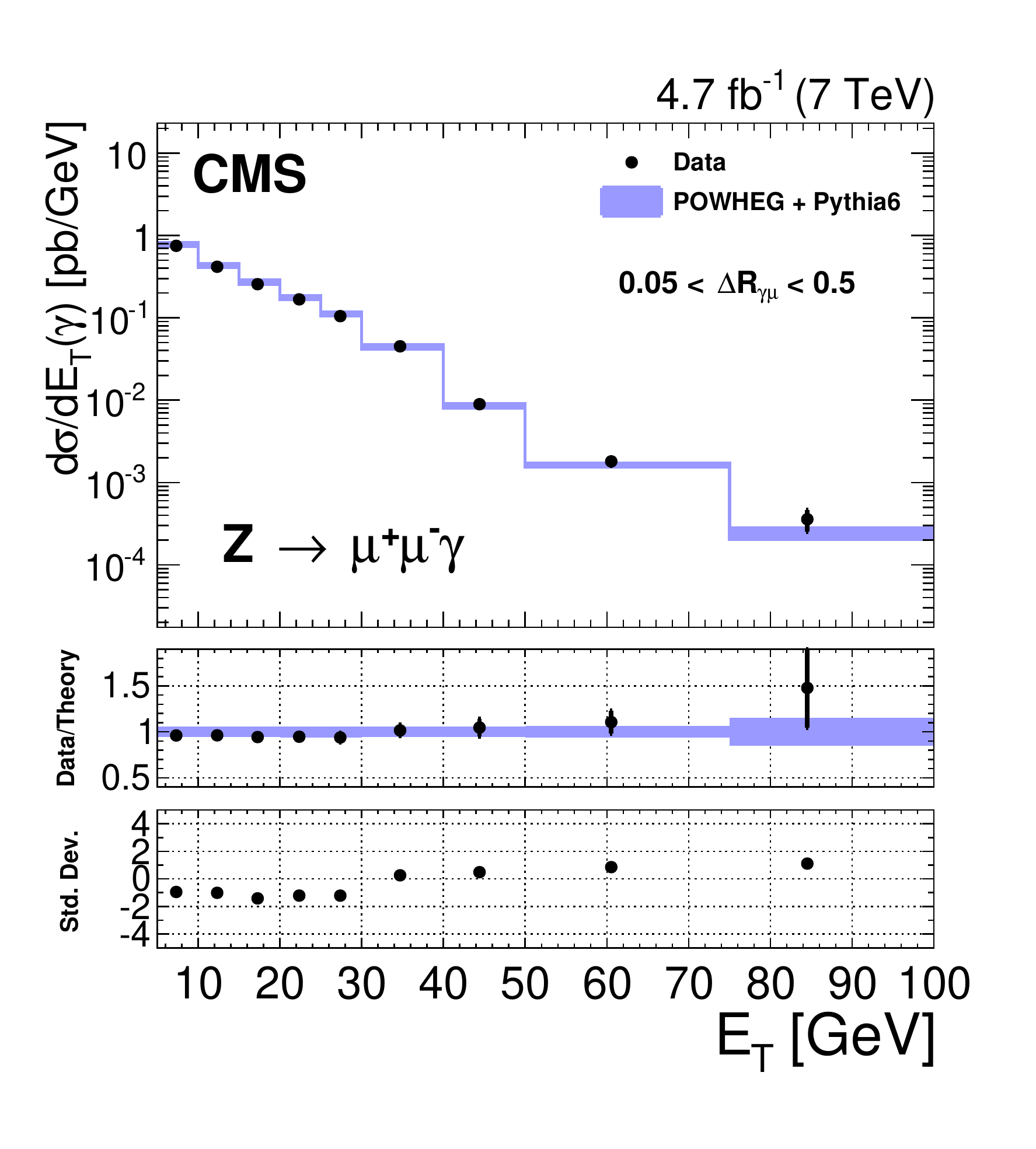}
\includegraphics[width=0.49\textwidth]{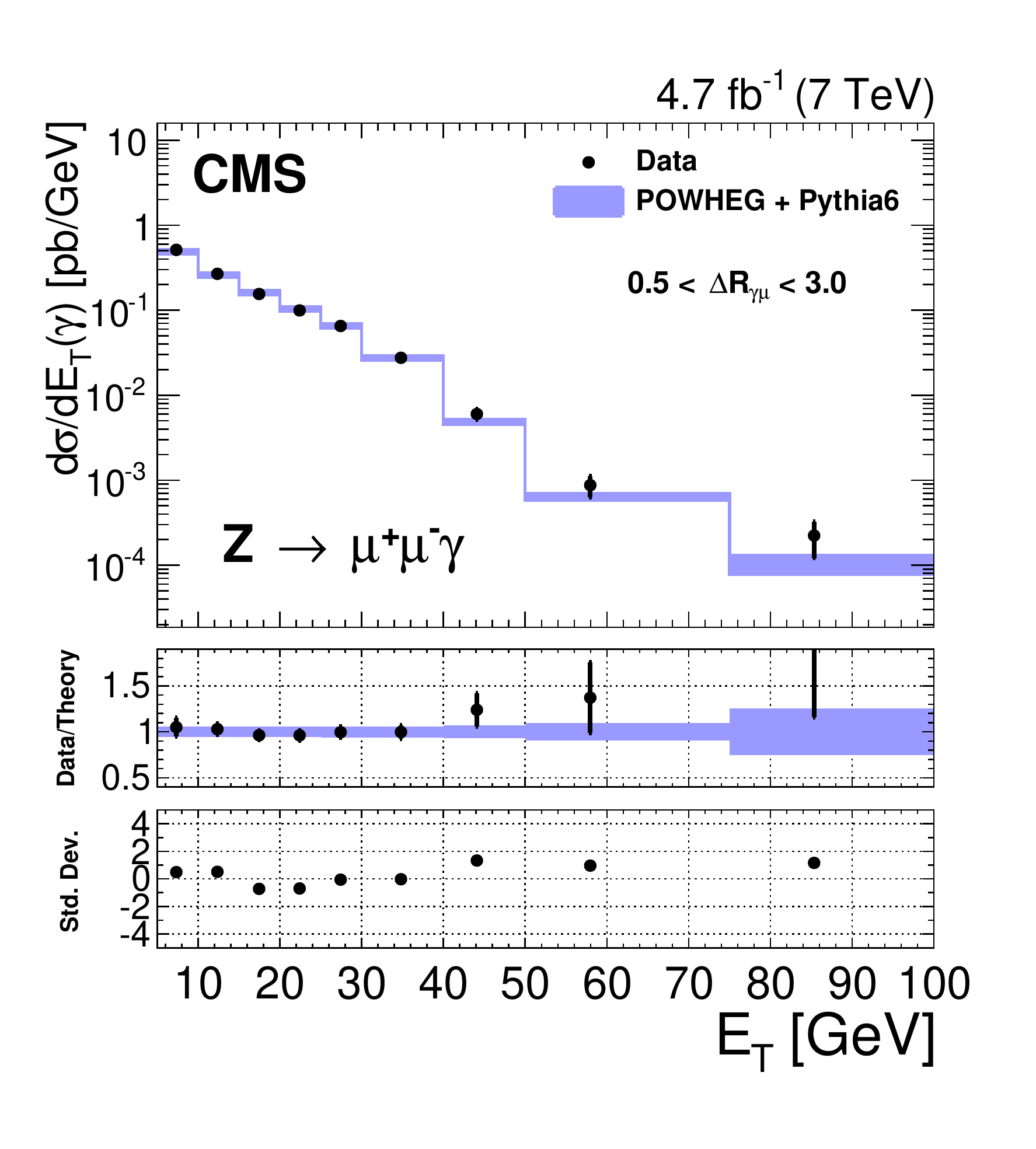}
\caption[.]{\label{fig:results2}
Measured differential cross sections $\DXDET$
for photons close to the muon ($0.05 < \DRmg \leq 0.5$,~\cmsLeft)
and far from the muon ($0.5 < \DRmg \leq 3$,~\cmsRight).
The dots with error bars represent the data, and the shaded
bands represent the \POWHEG{}+\PYTHIA calculation including theoretical
uncertainties. 
The central panels display the ratio of data to the MC expectation.
The lower panels show the standard deviations of the measurements
with respect to the calculation.
A bin-centering procedure has been applied.}
\end{figure*}

Energy spectra for photons closer to ($0.05 < \DRmg \leq 0.5$)
and farther from the muon ($0.5 < \DRmg \leq 3$) are
shown in Fig.~\ref{fig:results2}.  The rates for
photons with large $\DRmg$ and \ET are
also well reproduced.
The number of events with $30 < \Mmm < 87\GeV$ 
is about 18\% of the number with $60 < \Mmm < 120\GeV$.
Of the events with $30 < \Mmm < 87\GeV$, the fraction of events 
with at least one photon with
$\ET > 5\GeV$ and $0.05 < \DRmg \leq 0.5$
is~\FRACCLOSE, and with $0.5 < \DRmg \leq 3$ is~\FRACWIDE.
Photons with~$\DRmg > 1.2$ and~$\ET > 40\GeV$ constitute
a small fraction~\FRACRARE.

\begin{table*}[htbp]
\centering
\caption{\label{tab:dxdet}
Measured differential cross section $\DXDET$ in pb/\GeVns{}.
For the data values, the first uncertainty is statistical
and the second is systematic.  For the theory values,
the uncertainty combines statistical, PDF, and renormalization/factorization scale components.}
 \cmsTable{0.8\linewidth}{
 \begin{scotch}{lrr}

Kinematic requirement  [\GeVns{}]& \multicolumn{1}{c}{Data} & \multicolumn{1}{c}{\POWHEG{}+\PYTHIA} \\
\hline
\multicolumn{3}{c}{$0.05 < \DRmg \leq 3$} \\
\hline
    $5\phantom{0} < \ET \leq 10$ &  $\phantom{(}1.260 \pm 0.015 \pm 0.070 \phantom{)\times10^{-2}}$ & $\phantom{(} 1.270 \pm 0.075 \phantom{)\times10^{-2}}$ \\
    $10 < \ET \leq 15$ &  $\phantom{(}0.685 \pm 0.009 \pm 0.028 \phantom{)\times10^{-2}}$ & $\phantom{(} 0.694 \pm 0.040 \phantom{)\times10^{-2}}$ \\
    $15 < \ET \leq 20$ &  $\phantom{(}0.411 \pm 0.006 \pm 0.016 \phantom{)\times10^{-2}}$ & $\phantom{(} 0.433 \pm 0.025 \phantom{)\times10^{-2}}$ \\
    $20 < \ET \leq 25$ &  $\phantom{(}0.267 \pm 0.005 \pm 0.011 \phantom{)\times10^{-2}}$ & $\phantom{(} 0.280 \pm 0.017 \phantom{)\times10^{-2}}$ \\
    $25 < \ET \leq 30$ &  $\phantom{(}0.170 \pm 0.004 \pm 0.008 \phantom{)\times10^{-2}}$ & $\phantom{(} 0.177 \pm 0.011 \phantom{)\times10^{-2}}$ \\
    $30 < \ET \leq 40$ &  $(7.26 \pm\phantom{0}  0.19 \pm\phantom{0}  0.39)\times10^{-2} $ & $ (7.20 \pm\phantom{0}  0.42 )\times10^{-2} $ \\
    $40 < \ET \leq 50$ &  $(1.49 \pm\phantom{0}  0.09 \pm\phantom{0}  0.10)\times10^{-2} $ & $ (1.34 \pm\phantom{0}  0.08 )\times10^{-2} $ \\
    $50 < \ET \leq 75$ &  $(2.68 \pm\phantom{0}  0.26 \pm\phantom{0}  0.25)\times10^{-3} $ & $ (2.27 \pm\phantom{0}  0.14 )\times10^{-3} $ \\
    $75 < \ET \leq 100$ &  $(5.81 \pm\phantom{0}  1.16 \pm\phantom{0}  1.00)\times10^{-4} $ & $ (3.47 \pm\phantom{0}  0.32 )\times10^{-4} $ \\
\hline
\multicolumn{3}{c}{$0.05 < \DRmg \leq 0.5$} \\
\hline
    $5\phantom{0} < \ET \leq 10$ &  $\phantom{(}0.749 \pm 0.009 \pm 0.031 \phantom{)\times10^{-2}}$ & $\phantom{(} 0.779 \pm 0.045 \phantom{)\times10^{-2}}$ \\
    $10 < \ET \leq 15$ &  $\phantom{(}0.417 \pm 0.006 \pm 0.015 \phantom{)\times10^{-2}}$ & $\phantom{(} 0.433 \pm 0.025 \phantom{)\times10^{-2}}$ \\
    $15 < \ET \leq 20$ &  $\phantom{(}0.256 \pm 0.005 \pm 0.010 \phantom{)\times10^{-2}}$ & $\phantom{(} 0.272 \pm 0.016 \phantom{)\times10^{-2}}$ \\
    $20 < \ET \leq 25$ &  $\phantom{(}0.168 \pm 0.004 \pm 0.007 \phantom{)\times10^{-2}}$ & $\phantom{(} 0.177 \pm 0.011 \phantom{)\times10^{-2}}$ \\
    $25 < \ET \leq 30$ &  $\phantom{(}0.105 \pm 0.003 \pm 0.005 \phantom{)\times10^{-2}}$ & $\phantom{(} 0.112 \pm 0.007 \phantom{)\times10^{-2}}$ \\
    $30 < \ET \leq 40$ &  $(4.51 \pm\phantom{0}  0.14 \pm\phantom{0}  0.23)\times10^{-2} $ & $ (4.44 \pm\phantom{0}  0.26 )\times10^{-2} $ \\
    $40 < \ET \leq 50$ &  $(8.93 \pm\phantom{0}  0.65 \pm\phantom{0}  0.51)\times10^{-3} $ & $ (8.53 \pm\phantom{0}  0.50 )\times10^{-3} $ \\
    $50 < \ET \leq 75$ &  $(1.80 \pm\phantom{0}  0.18 \pm\phantom{0}  0.09)\times10^{-3} $ & $ (1.63 \pm\phantom{0}  0.10 )\times10^{-3} $ \\
    $75 < \ET \leq 100$ &  $(3.58 \pm\phantom{0}  0.98 \pm\phantom{0}  0.36)\times10^{-4} $ & $ (2.42 \pm\phantom{0}  0.37 )\times10^{-4} $ \\
\hline
\multicolumn{3}{c}{$0.5 < \DRmg \leq 3$} \\
\hline
    $5\phantom{0} < \ET \leq 10$ &  $\phantom{(}0.513 \pm 0.012 \pm 0.049 \phantom{)\times10^{-2}}$ & $\phantom{(} 0.489 \pm 0.028 \phantom{)\times10^{-2}}$ \\
    $10 < \ET \leq 15$ &  $\phantom{(}0.268 \pm 0.006 \pm 0.014 \phantom{)\times10^{-2}}$ & $\phantom{(} 0.260 \pm 0.015 \phantom{)\times10^{-2}}$ \\
    $15 < \ET \leq 20$ &  $\phantom{(}0.155 \pm 0.004 \pm 0.007 \phantom{)\times10^{-2}}$ & $\phantom{(} 0.161 \pm 0.010 \phantom{)\times10^{-2}}$ \\
    $20 < \ET \leq 25$ &  $(9.94 \pm\phantom{0}  0.33 \pm\phantom{0}  0.45)\times10^{-2} $ & $ (1.03 \pm\phantom{0}  0.06 )\times10^{-1} $ \\
    $25 < \ET \leq 30$ &  $(6.52 \pm\phantom{0}  0.26 \pm\phantom{0}  0.32)\times10^{-2} $ & $ (6.55 \pm\phantom{0}  0.39 )\times10^{-2} $ \\
    $30 < \ET \leq 40$ &  $(2.76 \pm\phantom{0}  0.12 \pm\phantom{0}  0.16)\times10^{-2} $ & $ (2.76 \pm\phantom{0}  0.17 )\times10^{-2} $ \\
    $40 < \ET \leq 50$ &  $(6.01 \pm\phantom{0}  0.67 \pm\phantom{0}  0.56)\times10^{-3} $ & $ (4.85 \pm\phantom{0}  0.33 )\times10^{-3} $ \\
    $50 < \ET \leq 75$ &  $(8.75 \pm\phantom{0}  1.86 \pm\phantom{0}  1.60)\times10^{-4} $ & $ (6.38 \pm\phantom{0}  0.60 )\times10^{-4} $ \\
    $75 < \ET \leq 100$ &  $(2.23 \pm\phantom{0}  0.63 \pm\phantom{0}  0.80)\times10^{-4} $ & $ (1.04 \pm\phantom{0}  0.27 )\times10^{-4} $ \\
\hline
\multicolumn{3}{c}{$0.05 < \DRmg \leq 3$ and $\QT < 10\GeV$} \\
\hline
    $5\phantom{0} < \ET \leq 10$ &  $\phantom{(}0.527 \pm 0.009 \pm 0.024 \phantom{)\times10^{-2}}$ & $\phantom{(} 0.535 \pm 0.033 \phantom{)\times10^{-2}}$ \\
    $10 < \ET \leq 15$ &  $\phantom{(}0.294 \pm 0.005 \pm 0.010 \phantom{)\times10^{-2}}$ & $\phantom{(} 0.296 \pm 0.018 \phantom{)\times10^{-2}}$ \\
    $15 < \ET \leq 20$ &  $\phantom{(}0.184 \pm 0.004 \pm 0.007 \phantom{)\times10^{-2}}$ & $\phantom{(} 0.191 \pm 0.012 \phantom{)\times10^{-2}}$ \\
    $20 < \ET \leq 25$ &  $\phantom{(}0.127 \pm 0.003 \pm 0.005 \phantom{)\times10^{-2}}$ & $\phantom{(} 0.129 \pm 0.008 \phantom{)\times10^{-2}}$ \\
    $25 < \ET \leq 30$ &  $(8.59 \pm\phantom{0}  0.28 \pm\phantom{0}  0.40)\times10^{-2} $ & $ (8.25 \pm\phantom{0}  0.54 )\times10^{-2} $ \\
    $30 < \ET \leq 40$ &  $(3.22 \pm\phantom{0}  0.12 \pm\phantom{0}  0.19)\times10^{-2} $ & $ (2.89 \pm\phantom{0}  0.18 )\times10^{-2} $ \\
    $40 < \ET \leq 50$ &  $(1.46 \pm\phantom{0}  0.27 \pm\phantom{0}  0.14)\times10^{-3} $ & $ (1.14 \pm\phantom{0}  0.12 )\times10^{-3} $ \\
    $50 < \ET \leq 75$ &  $(1.92 \pm\phantom{0}  0.67 \pm\phantom{0}  0.42)\times10^{-4} $ & $ (8.44 \pm\phantom{0}  1.60 )\times10^{-5} $ \\
    $75 < \ET \leq 100$ &  $(1.67 \pm\phantom{0}  2.10 \pm\phantom{0}  0.66)\times10^{-5} $ & $ (6.66 \pm\phantom{0}  5.13 )\times10^{-6} $ \\
\hline
\multicolumn{3}{c}{$0.05 < \DRmg \leq 3$ and $\QT > 50\GeV$} \\
\hline
    $5\phantom{0} < \ET \leq 10$ &  $\phantom{(}0.104 \pm 0.005 \pm 0.008 \phantom{)\times10^{-2}}$ & $\phantom{(} 0.095 \pm 0.005 \phantom{)\times10^{-2}}$ \\
    $10 < \ET \leq 15$ &  $(6.26 \pm\phantom{0}  0.28 \pm\phantom{0}  0.33)\times10^{-2} $ & $ (5.72 \pm\phantom{0}  0.31 )\times10^{-2} $ \\
    $15 < \ET \leq 20$ &  $(3.67 \pm\phantom{0}  0.20 \pm\phantom{0}  0.19)\times10^{-2} $ & $ (3.38 \pm\phantom{0}  0.18 )\times10^{-2} $ \\
    $20 < \ET \leq 25$ &  $(2.19 \pm\phantom{0}  0.15 \pm\phantom{0}  0.10)\times10^{-2} $ & $ (2.32 \pm\phantom{0}  0.13 )\times10^{-2} $ \\
    $25 < \ET \leq 30$ &  $(1.94 \pm\phantom{0}  0.14 \pm\phantom{0}  0.09)\times10^{-2} $ & $ (1.64 \pm\phantom{0}  0.10 )\times10^{-2} $ \\
    $30 < \ET \leq 40$ &  $(9.98 \pm\phantom{0}  0.71 \pm\phantom{0}  0.51)\times10^{-3} $ & $ (9.79 \pm\phantom{0}  0.55 )\times10^{-3} $ \\
    $40 < \ET \leq 50$ &  $(6.21 \pm\phantom{0}  0.55 \pm\phantom{0}  0.32)\times10^{-3} $ & $ (5.58 \pm\phantom{0}  0.33 )\times10^{-3} $ \\
    $50 < \ET \leq 75$ &  $(1.90 \pm\phantom{0}  0.20 \pm\phantom{0}  0.11)\times10^{-3} $ & $ (1.76 \pm\phantom{0}  0.11 )\times10^{-3} $ \\
    $75 < \ET \leq 100$ &  $(4.56 \pm\phantom{0}  0.95 \pm\phantom{0}  0.55)\times10^{-4} $ & $ (3.13 \pm\phantom{0}  0.30 )\times10^{-4} $ \\
 \end{scotch}
 }

 \end{table*}

\begin{table*}[htbp]
\begin{center}
\topcaption{\label{tab:dxddr}
Measured differential cross section $\DXDDR$ in pb.
For the data values, the first uncertainty is statistical
and the second is systematic.  For the theory values,
the uncertainty combines statistical, PDF, and renormalization/factorization scale components.}
{
 \begin{scotch}{lrr}
Kinematic & \multicolumn{1}{c}{Data} & \multicolumn{1}{c}{\POWHEG{}+\PYTHIA} \\
requirement & & \\
\hline
\multicolumn{3}{c}{$\ET > 5.0\GeV$} \\
\hline
    $0.05 < \DRmg \leq 0.1$ &  $\phantom{(}53.90 \pm 0.76 \pm 2.00 \phantom{)\times10^{-2}}$ & $\phantom{(} 56.60 \pm 3.26 \phantom{)\times10^{-2}}$ \\
    $0.1\phantom{0} < \DRmg \leq 0.15$ &  $\phantom{(}31.90 \pm 0.59 \pm 1.20 \phantom{)\times10^{-2}}$ & $\phantom{(} 33.20 \pm 1.96 \phantom{)\times10^{-2}}$ \\
    $0.15 < \DRmg \leq 0.3$ &  $\phantom{(}18.40 \pm 0.25 \pm 0.67 \phantom{)\times10^{-2}}$ & $\phantom{(} 19.00 \pm 1.10 \phantom{)\times10^{-2}}$ \\
    $0.3\phantom{0} < \DRmg \leq 0.5$ &  $\phantom{(}10.10 \pm 0.16 \pm 0.37 \phantom{)\times10^{-2}}$ & $\phantom{(} 10.50 \pm 0.59 \phantom{)\times10^{-2}}$ \\
    $0.5\phantom{0} < \DRmg \leq 0.8$ &  $\phantom{(}6.14 \pm 0.11 \pm 0.25 \phantom{)\times10^{-2}}$ & $\phantom{(} 6.29 \pm 0.37 \phantom{)\times10^{-2}}$ \\
    $0.8\phantom{0} < \DRmg \leq 1.2$ &  $\phantom{(}4.22 \pm 0.09 \pm 0.21 \phantom{)\times10^{-2}}$ & $\phantom{(} 4.10 \pm 0.24 \phantom{)\times10^{-2}}$ \\
    $1.2\phantom{0} < \DRmg \leq 1.6$ &  $\phantom{(}2.94 \pm 0.08 \pm 0.19 \phantom{)\times10^{-2}}$ & $\phantom{(} 2.91 \pm 0.17 \phantom{)\times10^{-2}}$ \\
    $1.6\phantom{0} < \DRmg \leq 2.0$ &  $\phantom{(}1.76 \pm 0.07 \pm 0.16 \phantom{)\times10^{-2}}$ & $\phantom{(} 1.79 \pm 0.11 \phantom{)\times10^{-2}}$ \\
    $2.0\phantom{0} < \DRmg \leq 3.0$ &  $\phantom{(}0.46 \pm 0.04 \pm 0.10 \phantom{)\times10^{-2}}$ & $\phantom{(} 0.33 \pm 0.02 \phantom{)\times10^{-2}}$ \\
\hline
\multicolumn{3}{c}{$\ET > \GeV$ and $\QT < 10~\GeV$} \\
\hline
    $0.05 < \DRmg \leq 0.1$ &  $\phantom{(}23.00 \pm 0.50 \pm 0.82 \phantom{)\times10^{-2}}$ & $\phantom{(} 24.40 \pm 1.53 \phantom{)\times10^{-2}}$ \\
    $0.1\phantom{0} < \DRmg \leq 0.15$ &  $\phantom{(}13.70 \pm 0.39 \pm 0.49 \phantom{)\times10^{-2}}$ & $\phantom{(} 14.20 \pm 0.88 \phantom{)\times10^{-2}}$ \\
    $0.15 < \DRmg \leq 0.3$ &  $\phantom{(}7.88 \pm 0.17 \pm 0.28 \phantom{)\times10^{-2}}$ & $\phantom{(} 8.21 \pm 0.51 \phantom{)\times10^{-2}}$ \\
    $0.3\phantom{0} < \DRmg \leq 0.5$ &  $\phantom{(}4.38 \pm 0.10 \pm 0.16 \phantom{)\times10^{-2}}$ & $\phantom{(} 4.48 \pm 0.28 \phantom{)\times10^{-2}}$ \\
    $0.5\phantom{0} < \DRmg \leq 0.8$ &  $\phantom{(}2.65 \pm 0.07 \pm 0.10 \phantom{)\times10^{-2}}$ & $\phantom{(} 2.67 \pm 0.17 \phantom{)\times10^{-2}}$ \\
    $0.8\phantom{0} < \DRmg \leq 1.2$ &  $\phantom{(}1.75 \pm 0.05 \pm 0.07 \phantom{)\times10^{-2}}$ & $\phantom{(} 1.75 \pm 0.11 \phantom{)\times10^{-2}}$ \\
    $1.2\phantom{0} < \DRmg \leq 1.6$ &  $\phantom{(}1.29 \pm 0.05 \pm 0.06 \phantom{)\times10^{-2}}$ & $\phantom{(} 1.25 \pm 0.08 \phantom{)\times10^{-2}}$ \\
    $1.6\phantom{0} < \DRmg \leq 2.0$ &  $\phantom{(}0.72 \pm 0.04 \pm 0.04 \phantom{)\times10^{-2}}$ & $\phantom{(} 0.79 \pm 0.05 \phantom{)\times10^{-2}}$ \\
    $2.0\phantom{0} < \DRmg \leq 3.0$ &  $\phantom{(}0.10 \pm 0.01 \pm 0.02 \phantom{)\times10^{-2}}$ & $\phantom{(} 0.09 \pm 0.01 \phantom{)\times10^{-2}}$ \\
\hline
\multicolumn{3}{c}{$\ET > 5.0\GeV$ and $\QT > 50\GeV$} \\
\hline
    $0.05 < \DRmg \leq 0.1$ &  $\phantom{(}4.94 \pm 0.23 \pm 0.21 \phantom{)\times10^{-2}}$ & $\phantom{(} 5.07 \pm 0.27 \phantom{)\times10^{-2}}$ \\
    $0.1\phantom{0} < \DRmg \leq 0.15$ &  $\phantom{(}2.97 \pm 0.18 \pm 0.11 \phantom{)\times10^{-2}}$ & $\phantom{(} 3.05 \pm 0.18 \phantom{)\times10^{-2}}$ \\
    $0.15 < \DRmg \leq 0.3$ &  $\phantom{(}1.71 \pm 0.08 \pm 0.07 \phantom{)\times10^{-2}}$ & $\phantom{(} 1.73 \pm 0.09 \phantom{)\times10^{-2}}$ \\
    $0.3\phantom{0} < \DRmg \leq 0.5$ &  $\phantom{(}0.95 \pm 0.05 \pm 0.04 \phantom{)\times10^{-2}}$ & $\phantom{(} 0.98 \pm 0.06 \phantom{)\times10^{-2}}$ \\
    $0.5\phantom{0} < \DRmg \leq 0.8$ &  $\phantom{(}0.62 \pm 0.04 \pm 0.03 \phantom{)\times10^{-2}}$ & $\phantom{(} 0.58 \pm 0.03 \phantom{)\times10^{-2}}$ \\
    $0.8\phantom{0} < \DRmg \leq 1.2$ &  $\phantom{(}0.44 \pm 0.03 \pm 0.03 \phantom{)\times10^{-2}}$ & $\phantom{(} 0.37 \pm 0.02 \phantom{)\times10^{-2}}$ \\
    $1.2\phantom{0} < \DRmg \leq 1.6$ &  $\phantom{(}0.22 \pm 0.03 \pm 0.02 \phantom{)\times10^{-2}}$ & $\phantom{(} 0.19 \pm 0.01 \phantom{)\times10^{-2}}$ \\
    $1.6\phantom{0} < \DRmg \leq 2.0$ &  $\phantom{(}0.13 \pm 0.02 \pm 0.02 \phantom{)\times10^{-2}}$ & $\phantom{(} 0.10 \pm 0.01 \phantom{)\times10^{-2}}$ \\
    $2.0\phantom{0} < \DRmg \leq 3.0$ &  $(8.45 \pm 1.38 \pm 2.10)\times10^{-2} $ & $ (3.62 \pm 0.24 )\times10^{-2} $ \\
 \end{scotch}
 }
 \end{center}
 \end{table*}

\begin{figure*}[htbp]
\includegraphics[width=0.49\textwidth]{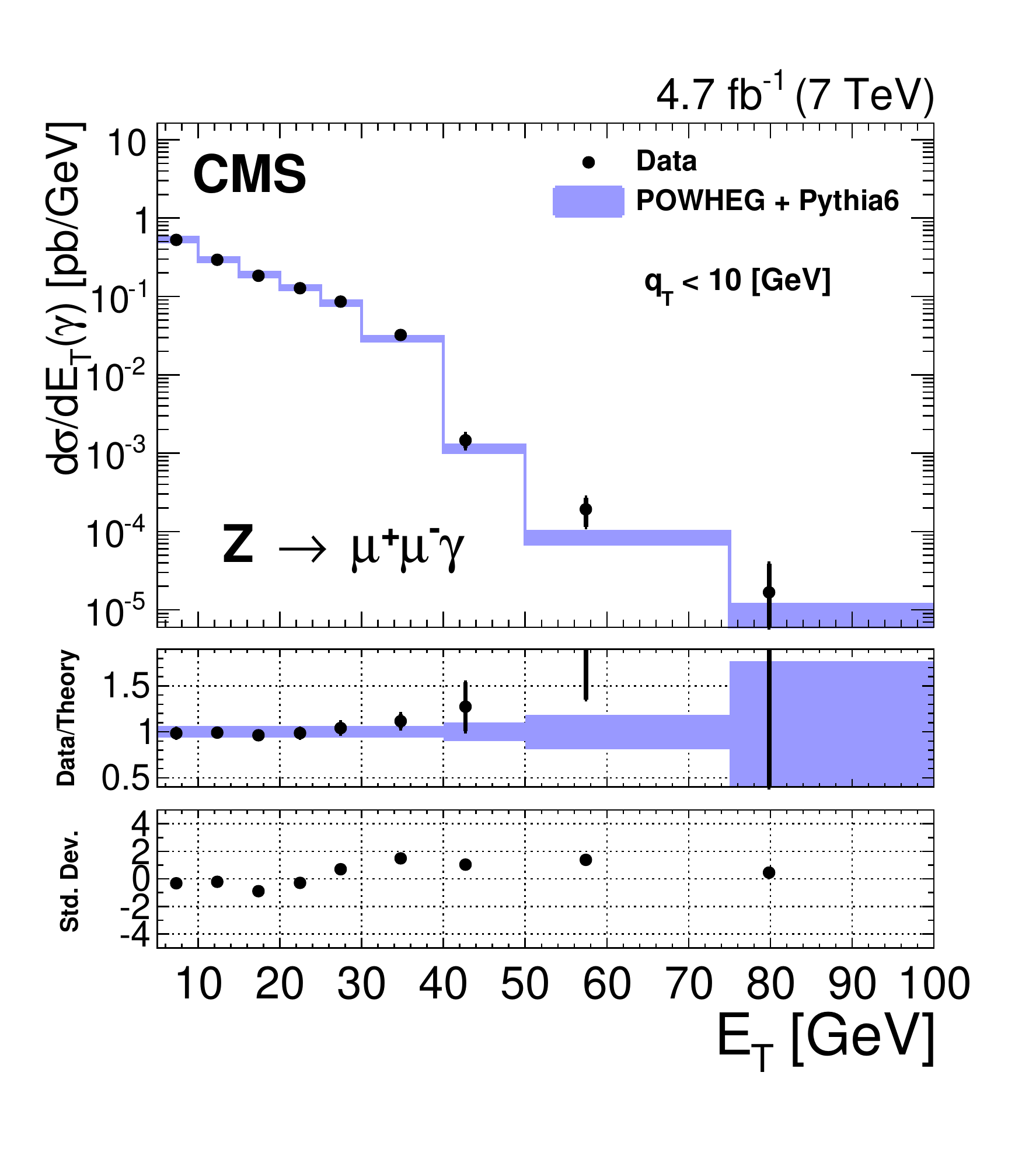}
\includegraphics[width=0.49\textwidth]{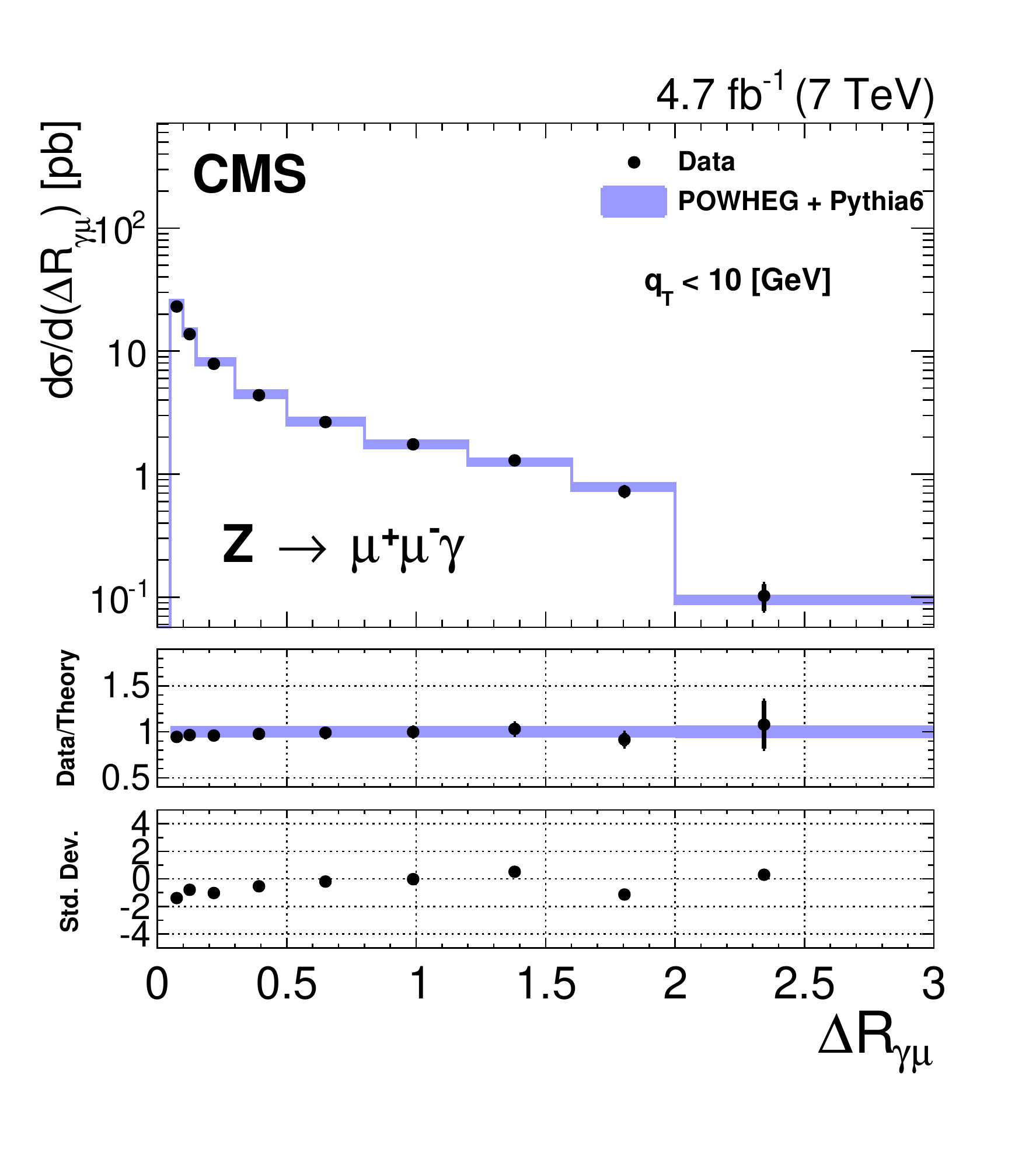}

\includegraphics[width=0.49\textwidth]{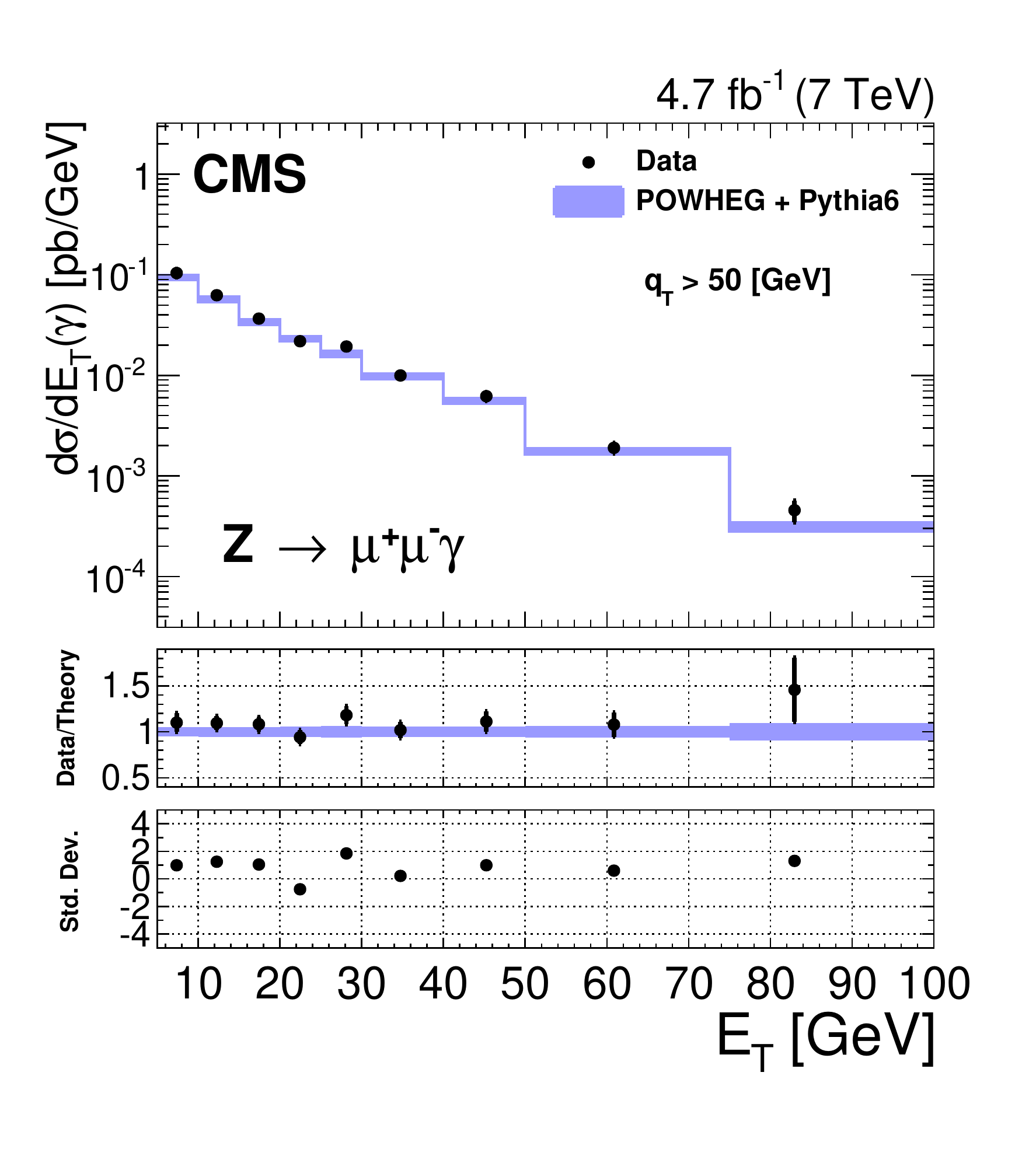}
\includegraphics[width=0.49\textwidth]{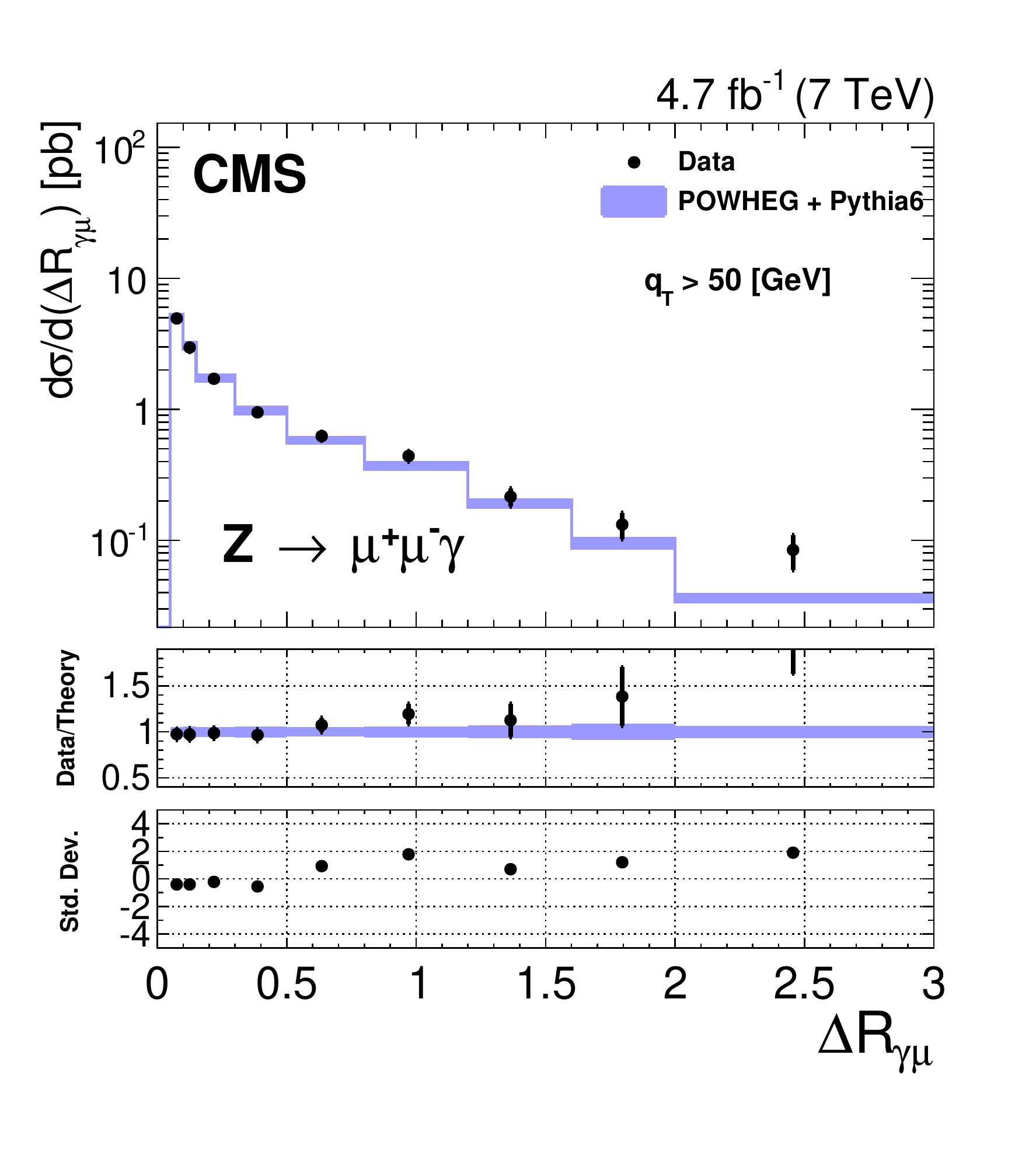}
\centering
\caption[.]{\label{fig:results3}
Measured differential cross sections $\DXDET$ and $\DXDDR$
for $\QT < 10\GeV$ (top row) and $\QT > 50\GeV$ (bottom row).
The dots with error bars represent the data, and the shaded
bands represent the \POWHEG{}+\PYTHIA calculation including theoretical
uncertainties.   
The central panels display the ratio of data to the MC expectation.
The lower panels show the standard deviations of the measurements
with respect to the calculation.
A bin-centering procedure has been applied.}
\end{figure*}

We define two subsamples of signal events, one with
the \Z boson transverse momentum
$\QT < 10\GeV$, and the other with $\QT > 50\GeV$.  The measured cross
sections shown in Fig.~\ref{fig:results3} demonstrate
rather different energy spectra for these two cases,
though $\DXDDR$ is basically the same.

\begin{figure}[htb]
\centering
\includegraphics[width=0.49\textwidth]{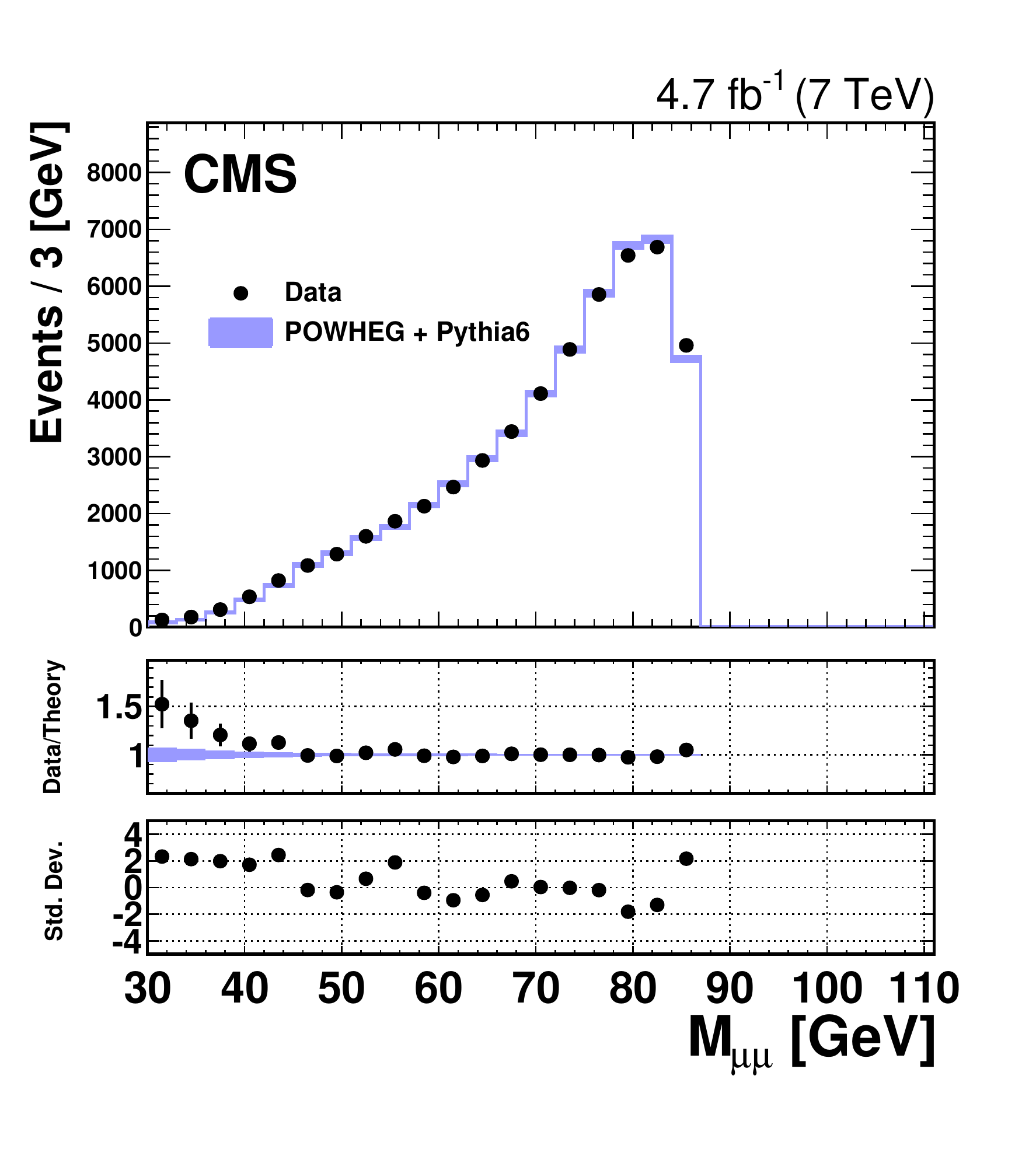}
\includegraphics[width=0.49\textwidth]{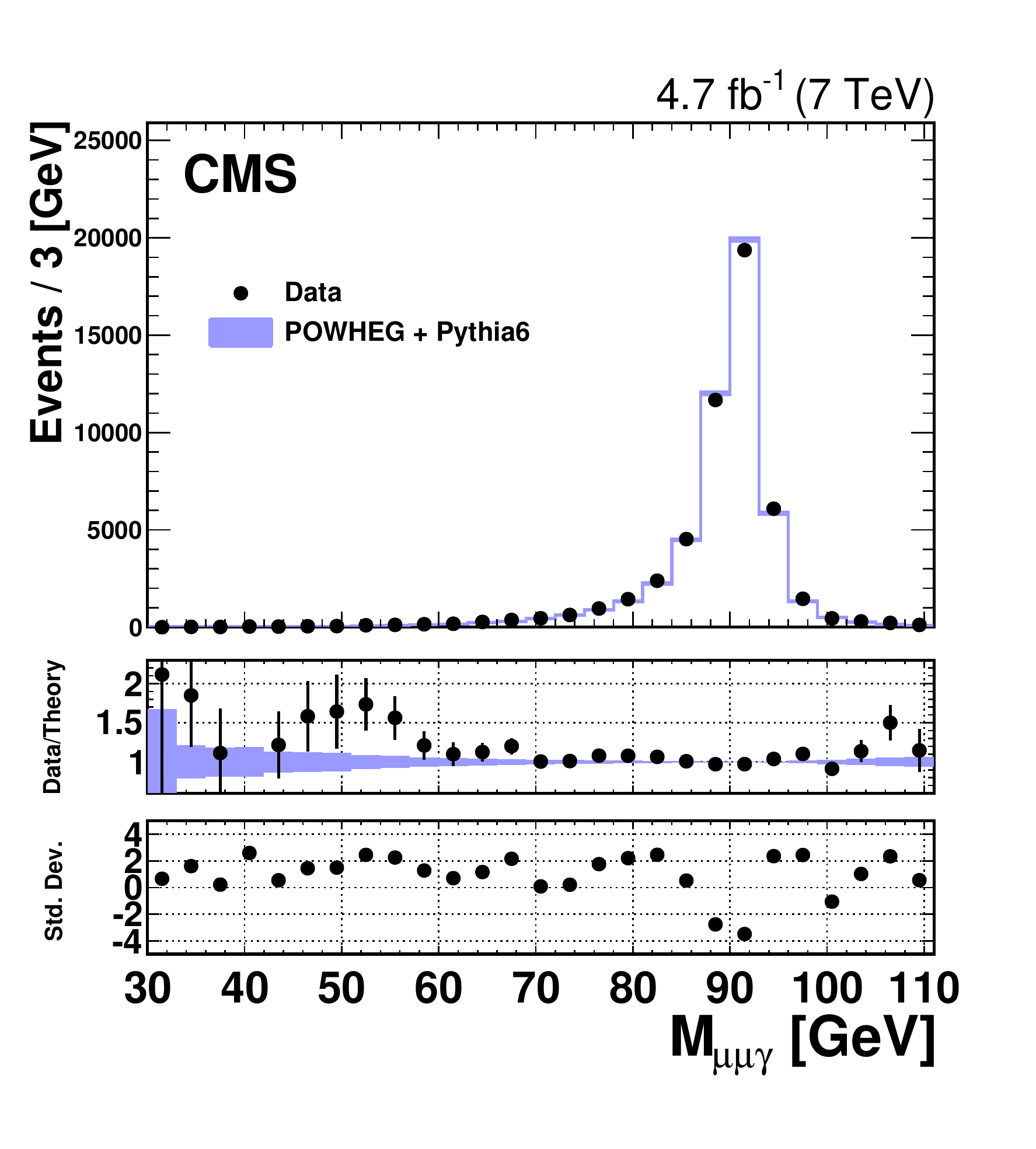}
\caption[.]{\label{fig:results4}
Distributions of the dimuon mass $\Mmm$~(\cmsLeft) and the
three-body mass $\Mmmg$~(\cmsRight).
The dots with error bars represent the data, and the shaded
bands represent the \POWHEG{}+\PYTHIA prediction.  
The central panels display the ratio of data to the MC expectation.
The lower panels show the standard deviations of the measurements
with respect to the calculation.
A bin-centering procedure has been applied.}
\end{figure}

As a final illustration of the nature of this event sample,
we present distributions of dimuon mass ($\Mmm$) and
the three-body mass ($\Mmmg$) in Fig.~\ref{fig:results4}.
The small increase in the ratio of data to theory for $\Mmm < 40\GeV$ reflects
the insufficient next-to-leading-order accuracy of the simulation; the kinematic requirements
on the muons induce a loss of acceptance that require higher-order
QCD corrections, as discussed in Ref.~\cite{DY}.
Although the masses of the dimuon pairs populate the tail
of the \Z~resonance (in fact they were selected this way),
the three-body mass distribution displays a nearly-symmetric
resonance peak at the mass of the \Z~boson, thereby confirming
the identity of these events as radiative decays
$\Z \to \mpmm\gamma$.

\section{Summary}
\label{sec:summary}

A study of final-state radiation in \Z~boson decays was presented.  This study
serves to test the simulation of events where mixed QED and QCD corrections are important.
The analysis was performed on a sample of $\Pp\Pp$ collision data at $\sqrt{s} = 7\TeV$ recorded in 2011 with the CMS detector
and corresponding to an integrated luminosity of \THELUMI.
Events with two oppositely charged muons and
an energetic, isolated photon were selected with only modest backgrounds.
The differential cross sections $\DXDET$ and $\DXDDR$ were measured for photons
within the fiducial and kinematic requirements specified in Table~\ref{tab:cuts},
and comparisons of $\DXDET$ for photons close to a muon and far from both muons were made.
In addition, the differential cross sections  $\DXDET$ and $\DXDDR$ were compared
for events with large and small transverse momentum of the Z boson, as computed from the
two muons and the photon.  Simulations based on \POWHEG{}+\PYTHIA reproduce the CMS data well, with discrepancies below 5\% for
$5 < \ET < 50\GeV$ and $0.05 < \DRmg \leq 2$
as quantified in Tables~\ref{tab:dxdet} and~\ref{tab:dxddr}.

\begin{acknowledgments}
We congratulate our colleagues in the CERN accelerator departments for the excellent performance of the LHC and thank the technical and administrative staffs at CERN and at other CMS institutes for their contributions to the success of the CMS effort. In addition, we gratefully acknowledge the computing centers and personnel of the Worldwide LHC Computing Grid for delivering so effectively the computing infrastructure essential to our analyses. Finally, we acknowledge the enduring support for the construction and operation of the LHC and the CMS detector provided by the following funding agencies: BMWFW and FWF (Austria); FNRS and FWO (Belgium); CNPq, CAPES, FAPERJ, and FAPESP (Brazil); MES (Bulgaria); CERN; CAS, MoST, and NSFC (China); COLCIENCIAS (Colombia); MSES and CSF (Croatia); RPF (Cyprus); MoER, ERC IUT and ERDF (Estonia); Academy of Finland, MEC, and HIP (Finland); CEA and CNRS/IN2P3 (France); BMBF, DFG, and HGF (Germany); GSRT (Greece); OTKA and NIH (Hungary); DAE and DST (India); IPM (Iran); SFI (Ireland); INFN (Italy); MSIP and NRF (Republic of Korea); LAS (Lithuania); MOE and UM (Malaysia); CINVESTAV, CONACYT, SEP, and UASLP-FAI (Mexico); MBIE (New Zealand); PAEC (Pakistan); MSHE and NSC (Poland); FCT (Portugal); JINR (Dubna); MON, RosAtom, RAS and RFBR (Russia); MESTD (Serbia); SEIDI and CPAN (Spain); Swiss Funding Agencies (Switzerland); MST (Taipei); ThEPCenter, IPST, STAR and NSTDA (Thailand); TUBITAK and TAEK (Turkey); NASU and SFFR (Ukraine); STFC (United Kingdom); DOE and NSF (USA).

Individuals have received support from the Marie-Curie program and the European Research Council and EPLANET (European Union); the Leventis Foundation; the A. P. Sloan Foundation; the Alexander von Humboldt Foundation; the Belgian Federal Science Policy Office; the Fonds pour la Formation \`a la Recherche dans l'Industrie et dans l'Agriculture (FRIA-Belgium); the Agentschap voor Innovatie door Wetenschap en Technologie (IWT-Belgium); the Ministry of Education, Youth and Sports (MEYS) of the Czech Republic; the Council of Science and Industrial Research, India; the HOMING PLUS program of Foundation for Polish Science, cofinanced from European Union, Regional Development Fund; the Compagnia di San Paolo (Torino); the Consorzio per la Fisica (Trieste); MIUR project 20108T4XTM (Italy); the Thalis and Aristeia programs cofinanced by EU-ESF and the Greek NSRF; and the National Priorities Research Program by Qatar National Research Fund.
\end{acknowledgments}
\bibliography{auto_generated}

\cleardoublepage \appendix\section{The CMS Collaboration \label{app:collab}}\begin{sloppypar}\hyphenpenalty=5000\widowpenalty=500\clubpenalty=5000\textbf{Yerevan Physics Institute,  Yerevan,  Armenia}\\*[0pt]
V.~Khachatryan, A.M.~Sirunyan, A.~Tumasyan
\vskip\cmsinstskip
\textbf{Institut f\"{u}r Hochenergiephysik der OeAW,  Wien,  Austria}\\*[0pt]
W.~Adam, T.~Bergauer, M.~Dragicevic, J.~Er\"{o}, M.~Friedl, R.~Fr\"{u}hwirth\cmsAuthorMark{1}, V.M.~Ghete, C.~Hartl, N.~H\"{o}rmann, J.~Hrubec, M.~Jeitler\cmsAuthorMark{1}, W.~Kiesenhofer, V.~Kn\"{u}nz, M.~Krammer\cmsAuthorMark{1}, I.~Kr\"{a}tschmer, D.~Liko, I.~Mikulec, D.~Rabady\cmsAuthorMark{2}, B.~Rahbaran, H.~Rohringer, R.~Sch\"{o}fbeck, J.~Strauss, W.~Treberer-Treberspurg, W.~Waltenberger, C.-E.~Wulz\cmsAuthorMark{1}
% ...
\vskip\cmsinstskip
% ...
1:~~Also at Vienna University of Technology, Vienna, Austria\\
2:~~Also at CERN, European Organization for Nuclear Research, Geneva, Switzerland\\
% ...

\end{sloppypar}
\end{document}